\documentclass[notitlepage,nofootinbib,superscriptaddress,aps,pre,twocolumn]{revtex4-1}
\pdfoutput=1
\usepackage[utf8]{inputenc}
\usepackage{amssymb}
\usepackage{physics}
\usepackage{graphicx}
\usepackage{color}
\usepackage{hyperref}
\usepackage{color}
\usepackage{bbm}
\usepackage{enumitem}
\newlist{myitemize}{enumerate}{10}
\setlist[myitemize]{label*=\arabic*.,nosep,leftmargin=*}
\newcommand{\be}{\begin{equation}}
\newcommand{\ee}{\end{equation}}
\newcommand{\bea}{\begin{eqnarray}}
\newcommand{\eea}{\end{eqnarray}}
\definecolor{colormp}{RGB}{0,105,10}
\begin{document}
\title{
A sprinkling of hybrid-signature discrete spacetimes in real-world networks
}
\author{Astrid Eichhorn}
\affiliation{CP3-Origins,  University  of  Southern  Denmark,  Campusvej  55,  DK-5230  Odense  M,  Denmark}
\author{Martin Pauly}
\affiliation{CP3-Origins,  University  of  Southern  Denmark,  Campusvej  55,  DK-5230  Odense  M,  Denmark}
\affiliation{Institut f\"ur Theoretische Physik, Ruprecht-Karls-Universit\"at Heidelberg, \\
Philosophenweg 16, 69120 Heidelberg, Germany}
\begin{abstract}
Many real-world networks are embedded into a space or spacetime.  The embedding space(time) constrains the properties of these real-world networks. We use the scale-dependent spectral dimension as a tool to probe whether real-world networks encode information on the dimensionality of the embedding space.
We find that spacetime networks which are inspired by quantum gravity and based on a hybrid signature, following the Minkowski metric at small spatial distance and the Euclidean metric at large spatial distance, provide a template relevant for real-world networks of small-world type, including a representation of the internet's architecture and biological neural networks.
\end{abstract}
\maketitle
\section{Introduction}
Networks occur in many different settings \cite{Strogatz,Albert:2002zz,2003SIAMR..45..167N,Boccaletti2006}, ranging from the fundamental structure of spacetime \cite{Bombelli:1987aa} to neural networks \cite{Bullmore2009,Bullmore2011,Sporns2011,Pessoa2014}. At a first glance, one might not expect these networks to be  similar to each other, given the very different settings they describe.
Yet at a second glance, these networks are related: in studies of spacetime, these are networks \emph{of} a geometry, in studies of neural networks, these are networks embedded \emph{in} a geometry.
Such spatial networks are constrained by the geometry in which they are embedded  \cite{barthelemy2011spatial}. For instance, networks of roads (mostly) lie  on top of two-dimensional landscapes. The landscape's dimensionality severely constrains the road networks. In turn, the road networks encode the dimensionality of the embedding landscapes, e.g., in the spectral dimension which can be measured by a diffusion process. The spectral dimension of road networks is typically around two, as we will calculate below. Road networks and the corresponding graphs are therefore simple and paradigmatic examples of how an embedding geometry determines properties of a network. Network geometry \cite{Bianconi:2015qpp,Mulder:2017oht,Boguna:2020gsy,Battiston2020} is a topic of active research and the importance of the spectral dimension in this context has been emphasized in \cite{Millan:2018dpa,Bianconi2020,2021JPCom...2a5001N}.
Many known networks exhibit small-world properties \cite{WS}. They contain ``shortcuts" that provide efficient connections between otherwise distant parts of the network. These ``shortcuts" can typically not be embedded into the original Euclidean embedding geometry at fixed dimensionality and topology.
To describe the existence of ``shortcuts'' in a geometric way, we propose to change how to measure distances in the underlying space; i.e., we propose to use a different metric.
A related idea has been put forward in the literature \cite{2009NatPh...5...74B,2010PhRvE..82c6106K}, see, e.g., \cite{hyperbolictrade,2015PhRvE..92c2812B} for applications, where it was suggested to measure distances according to a hyperbolic Euclidean metric. Points that lie far apart in a flat Euclidean metric\footnote{For the remainder of this paper, we refer to the flat Euclidean metric simply as the Euclidean metric.}, can be close in a such a hyperbolic metric.
Similarly, Lorentzian metrics can put points in close vicinity that lie far apart in a Euclidean metric.
Using the line element of the Minkowski metric
\be
\dd{s^2} = - \dd{t^2}+ \sum_{i=1}^{d-1} \dd{x_i^2},
\ee
a large spacelike distance $ \sum_{i=1}^{d-1} \dd{x_i^2}\gg 0$ can be compensated by a large timelike distance $ \dd{t^2} \gg 0$. All points on the light cone described by $t=r\equiv\sqrt{\sum x_i^2}$ lie at zero distance to each other.
As a consequence, any region of finite Minkowski distance extends infinitely far along the $t$ and $r$ directions.
Networks constructed according to the Minkowski metric hence contain ``shortcuts'' between nodes that are far apart according to the Euclidean metric. Therefore, the Minkowski metric is a promising candidate to capture a network's small-world properties.\\
The shortcuts also alter the spectral dimension which we use to test whether the Minkowski metric matters for real-world networks. \\
To perform this test, in this paper we first explore the spectral dimension for i) spacetime-networks arising with Euclidean, Minkowski and hybrid metric, ii) synthetic networks with small-world property, iii) networks based on data on road networks, internet architecture and brain connectivity in drosophila and mice.\\
The spectral dimension has been calculated for various graphs that are artificially generated, e.g., \cite{Durhuus:2006vk,Wheater:1998jb,Millan:2020plp}, including networks with small-world property \cite{Liu}, for graphs that encode the properties of certain real-world networks \cite{Bilke:2001nw} and for graphs that arise in approaches to quantum gravity \cite{Ambjorn:2005db,Benedetti:2009ge,Laiho:2011ya,Eichhorn:2013ova,Giasemidis:2013axa,Trugenberger:2015xma,Steinhaus:2018aav,Eichhorn:2019uct}. In quantum gravity, graphs exist for Euclidean as well as Lorentzian signature. The spectral dimension has already been analyzed for simplicial complexes, which naturally occur in particular quantum gravity approaches, but can be used more generally in the study of complex systems \cite{Reitz:2020mmz,Bianconi:2019qan}.
A systematic comparison of the diffusion processes on real-world networks with Lorentzian quantum gravity networks \cite{Eichhorn:2013ova} has not yet been performed. In this paper, we aim at closing this gap. In addition, we define a hybrid metric that is Minkowski at small (Euclidean, i.e., spatial) distances and Euclidean at large (Euclidean) distances and compare the spectral dimension of hybrid-signature spacetimes to that of real-world networks with small-world properties.
\section{Diffusion processes and scale-dependent spectral dimension}
The diffusion equation in the continuum determines the probability density at the point $x$, $P(x,x',\sigma)$ as a function of diffusion time $\sigma$:
\be
\left(\partial_{\sigma} - \nabla^2\right)P(x,x',\sigma)=0
.\ee
We have chosen units in which the diffusion constant is set to one. To solve the diffusion equation, we set the initial condition
\be
P(x,x',0) = \delta^d(x-x'),
\ee
describing a diffusion process that starts at $\sigma=0$ at point $x'$.
The resulting solution in the absence of curvature is
\be
P(x,x',\sigma) =\left(4\pi \sigma \right)^{-d/2} \exp\left(- \frac{(x-x')^2}{4\sigma}\right).
\ee
The return probability $P(x',x',\sigma)$ contains the information on the spectral dimension $d_s$:
\be
d_s =\underset{\sigma \rightarrow 0}{\rm lim}\left(-2 \frac{\partial \ln P(x',x',\sigma)}{\partial \ln \sigma}\right).\label{eq:dscont}
\ee
It agrees with the topological dimension, $d_s=d$.\\
The spectral dimension generalizes to networks. In contrast to the continuum diffusion process, the diffusion process or random walk is discrete on a network. The diffusion time is measured in integer values and denotes the number of discrete steps a random walker has taken. In each step, the random walker chooses among the edges of its present node with a weight assigned to each of the edges. In the simplest case the random walk progresses to any of the neighbouring nodes with equal probability.\\
For the first few steps of this random walk, the return probability oscillates between zero and non-zero.  On a regular lattice, this oscillation persists to arbitrarily large diffusion times, because it always takes an even number of steps to return to the starting point. On less regular networks, the oscillations disappear at larger diffusion times, because paths of even as well as uneven numbers of steps lead back to the starting point.
The oscillations can be smoothed out by implementing a finite probability $1-\delta$, ($\delta \in (0,1)$) to remain at the same node in each step. Choosing $\delta<1$ amounts to including the node itself among the set of its nearest neighbors. We choose $\delta=1/2$, motivated by the results of our numerical experiments.\\
From such a discrete random walk, we extract the spectral dimension $d_{\rm spec}$ of the network as a function of $\sigma$
\be
d_{\rm spec}(\sigma) = - 2 \frac{\sigma} {P(x',x',\sigma)} \frac{\Delta P}{\Delta \sigma}(x', \sigma).
\ee
with
\be
\frac{\Delta P}{\Delta \sigma}(x', \sigma) = \frac{P(x',x',\sigma +1) - P(x',x',\sigma-1)}{2}
\ee
The full scale-dependence of the spectral dimension contains information on the underlying network, as has been emphasized in the context of quantum gravity, see, e.g.,  \cite{Ambjorn:2005db,Reuter:2011ah,Calcagni:2013sca,Calcagni:2013vsa}.\\
Generically, the network spectral dimension features the following regimes for a finite network:\\
i) at small $\sigma$, the diffusion process probes the local neighborhood of the starting point $x'$. Thus, at small $\sigma$, $d_{\rm spec}(\sigma)$ encodes information on the local connectivity.\\
ii) At intermediate $\sigma$, $d_{\rm spec}$ may exhibit a plateau regime, i.e., $d_{\rm spec}(\sigma) \approx {\rm const} \approx d$, for such networks which can be embedded into a $d$-dimensional space, see, e.g., \cite{Ambjorn:2005db,Durhuus:2009sm,Benedetti:2009ge,Giasemidis:2012rf,Eichhorn:2019uct} in the context of quantum gravity.\\
iii) at large $\sigma$, equilibration sets in, i.e., $\partial_{\sigma}P(x,x',\sigma)\rightarrow 0$ due to the finiteness of the network and thus $d_{\rm spec} \rightarrow 0$ as a consequence of finite-size effects.\\
For infinitely extended graphs, the third regime vanishes, \cite{Durhuus:2009zz}, and a spectral dimension can be extracted in the limit $\sigma \rightarrow \infty$ that does not suffer from the discretization artefacts of i).
\section{Synthetic networks}
\begin{figure}[!t]
\includegraphics[width=\linewidth]{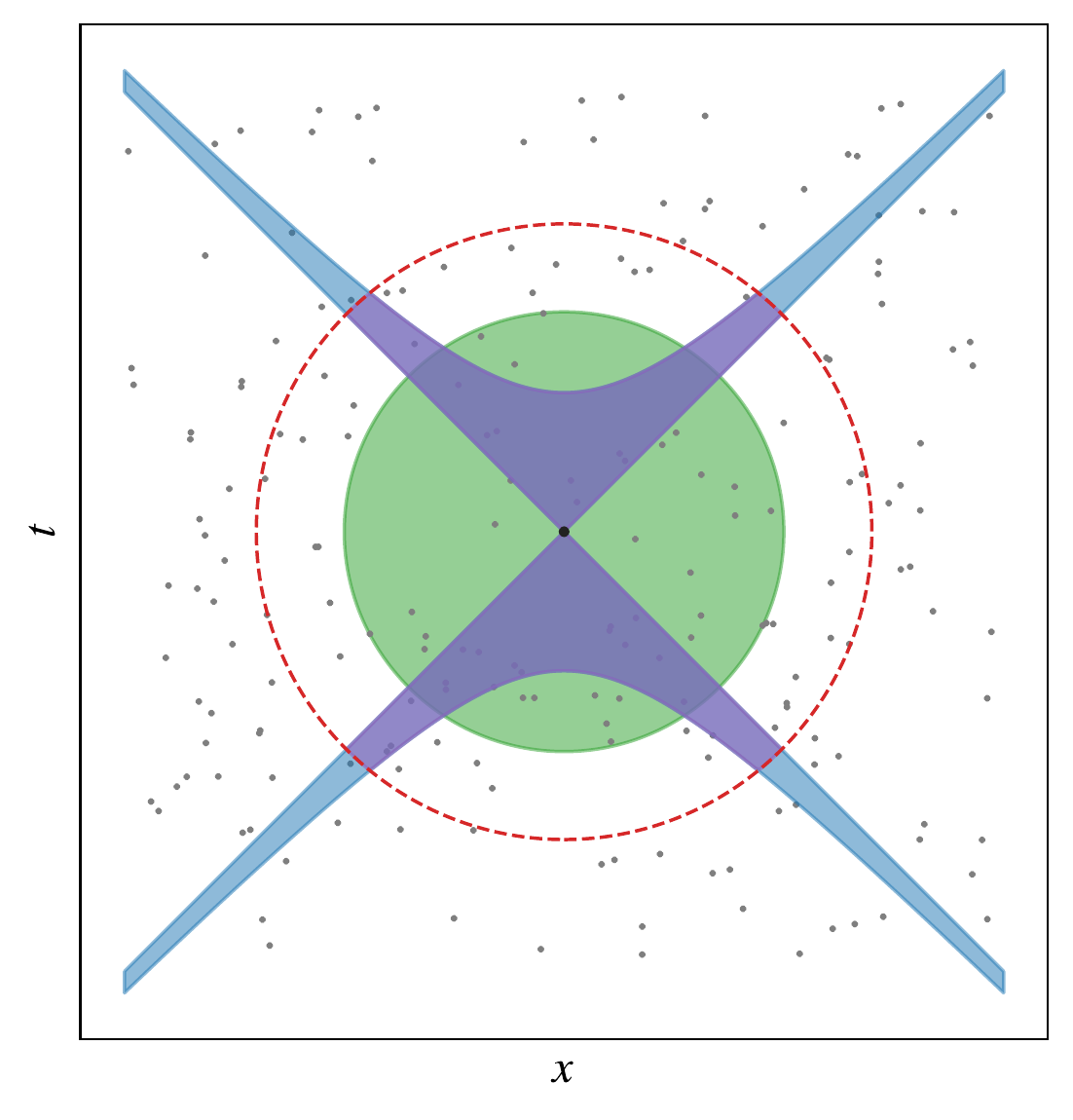}
\caption{\label{fig:sprinklingmetrics} We show a random distribution of points that are the nodes in a network. Adding edges according to the Euclidean metric generates a finite number of nearest neighbors, lying inside the green circle for the node in the center. Adding edges according to the Minkowski metric one obtains a number of nearest neighbors that grows with the size of the network, lying in the blue region between the light cone and the hyperboloid. Adding edges according to a hybrid which uses the Lorentzian metric inside a region delineated using the Euclidean metric (red dashed circle) cuts off the infinitely far extending region between light cone and hyperboloid such that one obtains finitely many nearest neighbors (inside the purple region).}
\end{figure}

Synthetic networks  provide templates for specific properties of networks, such as the network spectral dimension $d_{\rm spec}$. Here, we calculate the $\sigma$-dependence of $d_{\rm spec}$ for two important types of synthetic networks. \\
The first type are spacetime networks. In these networks, distances and nearest neighbors of a node are determined by a spacetime metric.  We investigate both Euclidean and Lorentzian metric signature. The main difference between the two is that Euclidean networks are local, in that nodes have a small and bounded number of nearest neighbors, whereas Lorentzian networks are nonlocal, in that nodes have a large and even unbounded number of nearest neighbors, cf.~Fig.~\ref{fig:sprinklingmetrics}. We also introduce a network that is based on a hybrid metric: below a Euclidean cutoff (i.e., a cutoff on the spatial distance), distances are measured in Lorentzian signature, above the cutoff they are measured in a Euclidean metric.
This hybrid network is semi-local, with a large but finite number of nearest neighbors.\\
The second type is the Watts-Strogatz model \cite{WS} see \cite{2000cond.mat..1118N,Strogatz} for reviews, which is the paradigmatic example of a network that exhibits the small-world property.
Watts-Strogatz models can be obtained starting from Euclidean spacetime networks. We will investigate the transition between the two in different embedding dimensions $d$. \\
For both types of networks, we investigate the network spectral dimension. This provides us with templates for the analysis of the spectral dimension in selected real-world networks with and without small-world property in Sec.~\ref{Sec:realnetworks}.
\subsection{Networks encoding Euclidean and Minkowski spacetime}
We generate a $d$-dimensional geometric network by selecting the $d$ coordinates of spacetime points which constitute the nodes of our networks.
Edges connect a node to its nearest neighbors according to the Euclidean or Minkowski metric. \\
In a Euclidean metric, all points at fixed distance to a central point lie on the surface of a $d$-sphere. Accordingly, the nearest neighbors of a node lie inside a $d$-sphere.\\
In  a Minkowski metric, all points at fixed distance to a central point lie on a hyperboloid. Accordingly, the nearest neighbors of a node lie just in between the $d$-hyperboloid and the lightcone, cf.~Fig.~\ref{fig:sprinklingmetrics}.
Assuming a finite density of points, the resulting Minkowski network has a degree that grows with the total volume of the Minkowski spacetime, whereas the
Euclidean network has finite and constant degree. This difference in the degree causes critical differences between the spectral dimensions in the two cases. \\
Points at small Minkowskian distance can lie at either large or small Euclidean distance (the converse is not true). Accordingly,  we call networks based on a Euclidean metric local, whereas we call networks based on the Minkowski metric non-local. Non-locality changes the spectral dimension drastically. In addition, it underlies the onset of small-world effects and is a prerequisite for those effects.\\
We will use a different distribution of coordinates for the Euclidean than for the Minkowski case for practical reasons.
The corresponding $d$-dimensional flat continuous space or spacetime, endowed with the Euclidean or Minkowski metric, features a global $SO(d)$ or $SO(d-1,1)$ symmetry, i.e., rotational symmetry and Lorentz symmetry. These symmetries are broken by a selection of the nodes' coordinates and one can choose whether or not to preserve a subgroup. A regular grid preserves a $\mathbbm{Z}_d$ symmetry of the $SO(d)$ symmetry, and a $\mathbb{Z}_{d-1}$ symmetry of the  $SO(d-1,1)$ symmetry, but breaks boost symmetry completely. In contrast, a random distribution of points preserves the $SO(d)$ or $SO(d-1,1)$ symmetry in a statistical sense, i.e., when averaging over many realizations of the random distribution. With a quantum-gravity motivation in mind, the statistical realization of the symmetries is preferable, because of strong experimental constraints on the explicit breaking of Lorentz symmetry \cite{Mattingly:2005re}. A non-uniform distance between nodes  is also preferable for many real-world networks because it allows to assign non-uniform weights to the nearest neighbors, which fits the network structure in many real-world-networks, see, e.g., \cite{Bassett3}.
Therefore we work with a random distribution (called a sprinkling \cite{Dowker:2013dog,Surya:2019ndm}) in the Minkowski case.\\
We choose a regular grid for the Euclidean case for simplicity. The difference between a regular grid and a random distribution is immaterial for the spectral dimension, because the average number of nearest neighbors is exactly $2^d$ for regular grids and on average $2^d$ for random distributions, when a Euclidean metric is used.
\subsection{Networks for Euclidean spacetime, their small degree and their spectral dimension}\label{sec:grid}
\begin{figure}[!t]
\includegraphics[width=\linewidth]{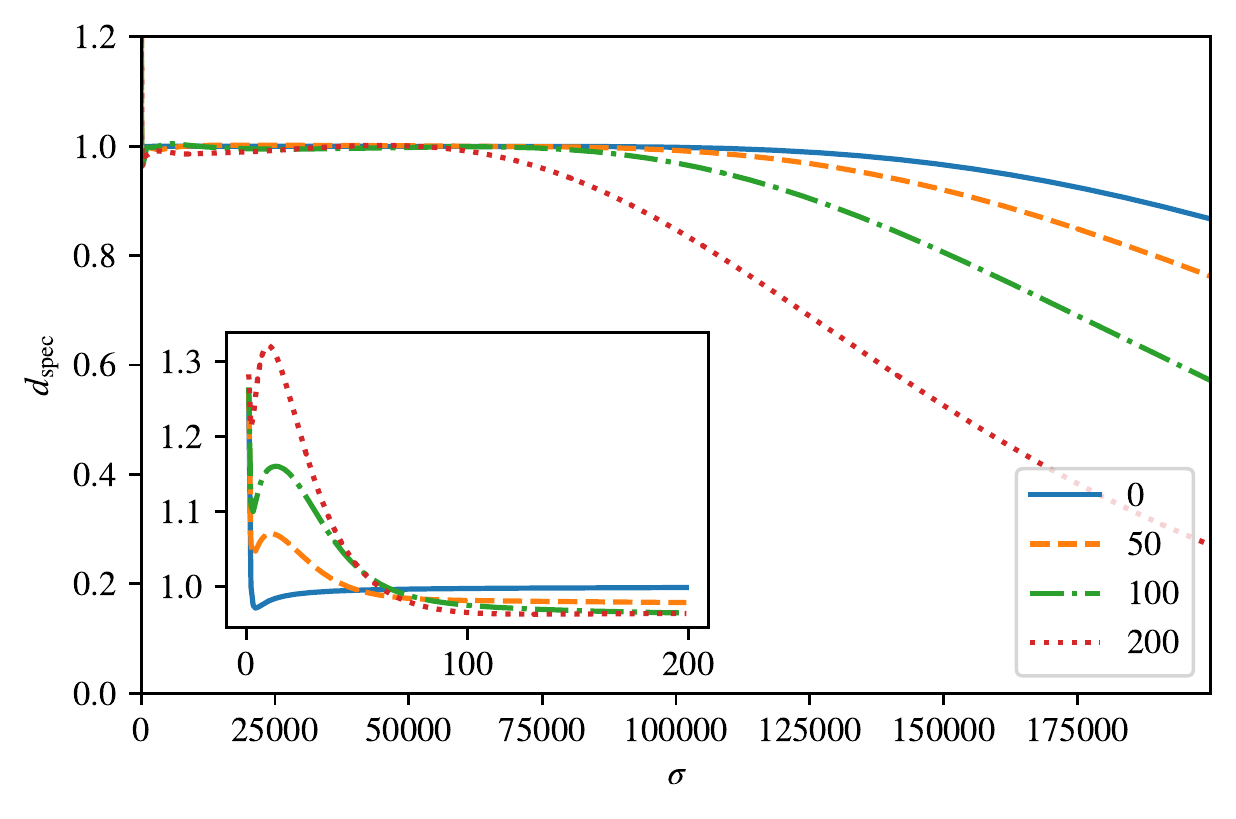}
\includegraphics[width=\linewidth]{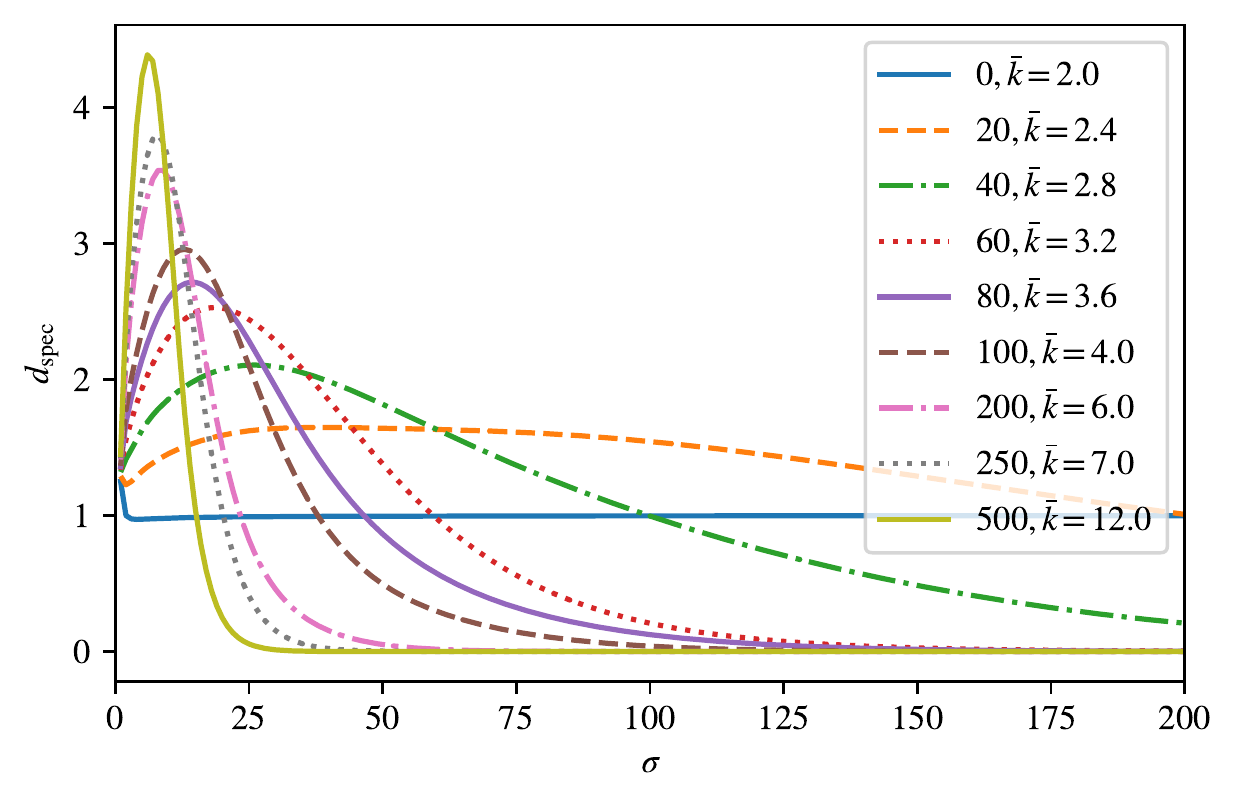}
\caption{\label{fig:ring}
Upper panel:
We show the spectral dimension for a ring with 1000 nodes. Additional connections are added randomly between previously unconnected pairs of nodes, if they are shorter than $l_\text{max}=10$.\\
Lower panel: We show the spectral dimension for a ring with 100 nodes, each of which is connected to two nearest neighbors, so that the graph can be embedded in a circle. Additional connections are added randomly between previously unconnected pairs of nodes, triggering a transition to a small-world network. The average degree $\bar{k}$ is also indicated.
}
\end{figure}

For a regular, infinitely extended grid embedded in $d$ dimensions with edges connecting neighbors according to the Euclidean metric, the network spectral dimension $d_{\rm spec}$ is lower \footnote{Note that this property is not always fulfilled for very small $\sigma$ in the examples we present. This is a consequence of how we discretize the derivative  and introduce the parameter $\delta$.} than the topological dimension $d$. The two dimensions agree only asymptotically, i.e., in the limit $\sigma \rightarrow \infty$, if the networks has infinitely many nodes. For a finite, but large enough, network, $d_{\rm spec}\lesssim d$ holds along a plateau at intermediate $\sigma$, before equilibration sets in and $d_{\rm spec} \rightarrow 0$.
As a simple first example, we consider a ring with 1000 nodes embedded in the circle, $S^1$. Each node is connected to its two nearest neighbors. The spectral dimension $d_{\rm spec}$ lies below 1 at small $\sigma$. This is a well-known consequence of the discretization of the continuous one-dimensional space. The discretization effect disappears, once $\sigma$ is large enough. At these intermediate values of $\sigma$, a plateau at $d_{\rm spec}\approx 1$ is reached, reflecting the embedding geometry, cf.~upper panel of Fig.~\ref{fig:ring}. At large $\sigma$, the diffusion process equilibrates, because the graph has a finite number of nodes. Thus, the spectral dimension approaches zero.\\
Next, we add non-locality to the network.\\
First, we only allow semi-local edges and randomly add new edges between two nodes whose original network distance is shorter than $l_{\rm max}=10$. \\
As a first consequence of the added edges, $d_{\rm spec}>d$ at small $\sigma$. This happens because
the neighborhood of a node, i.e., the set of nodes it shares an edge with, now has a finite probability to contain more than two nodes.
Accordingly, it resembles the local neighborhood of a node in a regular grid embedded in higher-dimensional space. \\
As a second consequence, the transition between the plateau and the drop-off occurs earlier. This happens because the effective size of the ring is smaller and thus equilibration occurs earlier.\\
Next, we drop the semi-locality restriction on the newly added edges. We add edges to the network between randomly selected pairs of previously unconnected nodes, see also \cite{newman1999renormalization,newman1999scaling}. This results in both semi-local connections that provide ``shortcuts" over short distances and non-local connections that provide ``shortcuts" over long distances.
The effect of these additional connections is shown in the lower panel of Fig.~\ref{fig:ring}, which features a ring with 100 nodes that is embedded in a one-dimensional space.
The addition of new edges results in a sharp increase of the spectral dimension at low diffusion times, and a peak at $d_{\rm peak}>1$, the height of which depends on the average degree $\bar{k}$.
Simultaneously, the onset of equilibration occurs earlier and thus the plateau at $d_{\rm spec}=1$ no longer exists if the number of extra connections is too large compared to the total number of nodes.\\
Comparing the upper and lower panel of Fig.~\ref{fig:ring}, we observe a dramatic impact of non-local connections. In contrast, added semi-local connections impact the spectral dimension much less: they speed up the onset of equilibration, but do not significantly change the spectral dimension at intermediate $\sigma$.\\
The local network does not feature small-world properties \cite{newman1999renormalization,newman1999scaling}: the average distance grows faster than logarithmically with the number of nodes $N$. For a grid embedded in $d$-dimensional space, the average distance between any two nodes is $\sim N^{\frac{1}{d}}$. It is evident that the addition of non-local edges dramatically impacts that scaling. In the limit of the fully connected graph, the average distance is 1 and does not depend on $N$. In between the fully connected graph and the Euclidean local network with scaling $\sim N^{\frac{1}{d}}$, a logarithmic dependence on $N$ can be achieved, which signals the onset of small-world properties.\\
The transition between a ring and a nonlocal model highlights two key points that are important in the following:\\
i) a network that can be embedded in a space of topological dimension $d$ with edges drawn according to the Euclidean metric exhibits a plateau in the spectral dimension at $d_{\rm spec} = d$ and approaches $d$ from below;\\
ii) additional connections drive up the spectral dimension at small $\sigma$, and have the largest impact on the spectral dimension when they are non-local, i.e., connect nodes that lie at large Euclidean distance.
\subsection{Networks for Minkowski spacetime, their unbounded degree and their spectral dimension}
\label{sec:causal_sets_spectral_dimension}
\begin{figure}
\includegraphics[width=\linewidth]{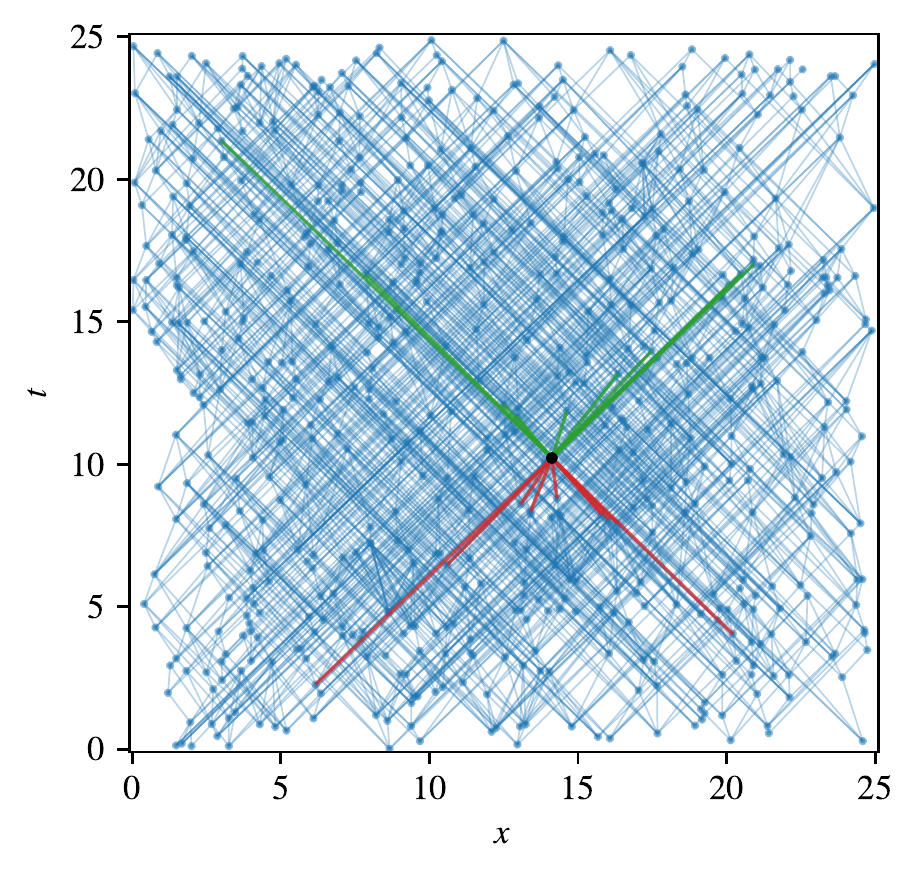}
\caption{\label{fig:sprinkling} We show a sprinkling into a finite region in 1+1 dimensional Minkowski spacetime. We highlight the direct future (green) and direct past (red) of the event that is represented by the larger black dot. The nearest neighbors approximate a hyperboloid.}
\end{figure}

\begin{figure}
\includegraphics[width=\linewidth]{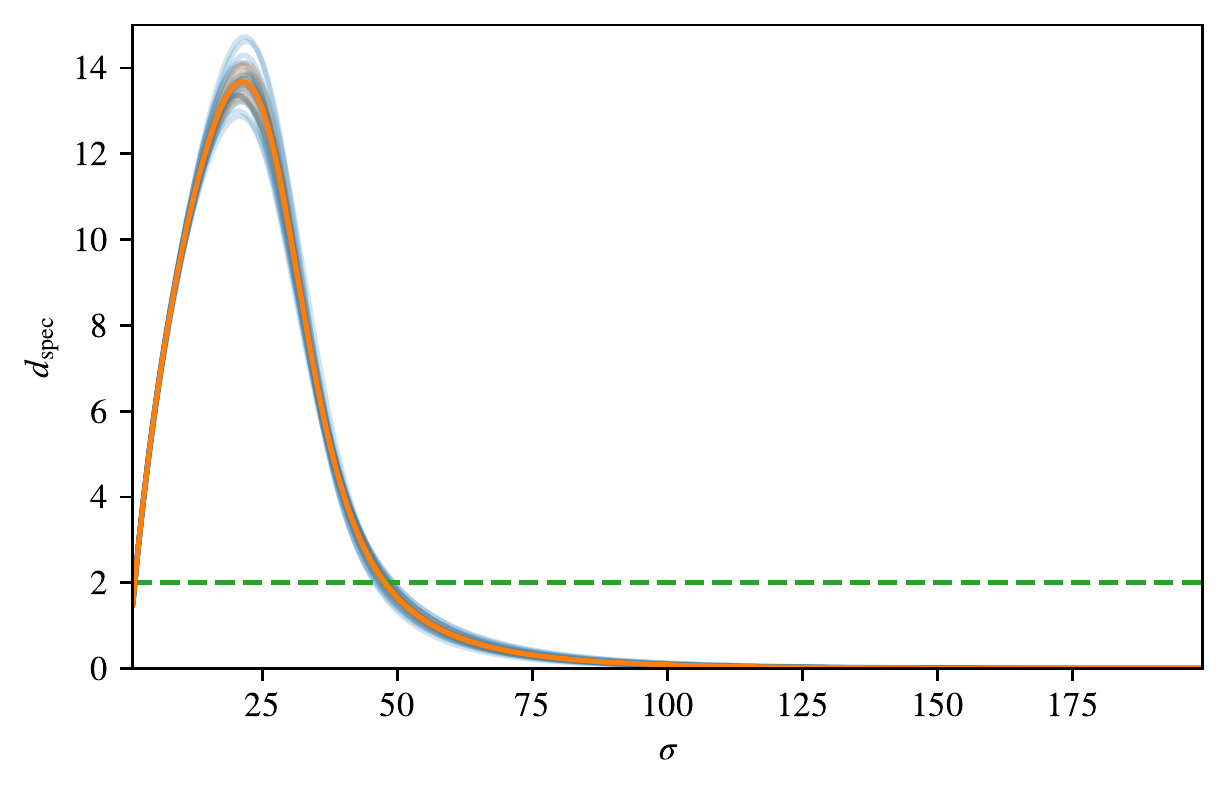}
\caption{\label{fig:diffcsnocutoff} We show the spectral dimension of a causal set that embeds into 2-dimensional Minkowski spacetime. The green dashed line marks $d_\text{spec}= 2$.}
\end{figure}

\begin{figure}[!t]
\includegraphics[width=\linewidth]{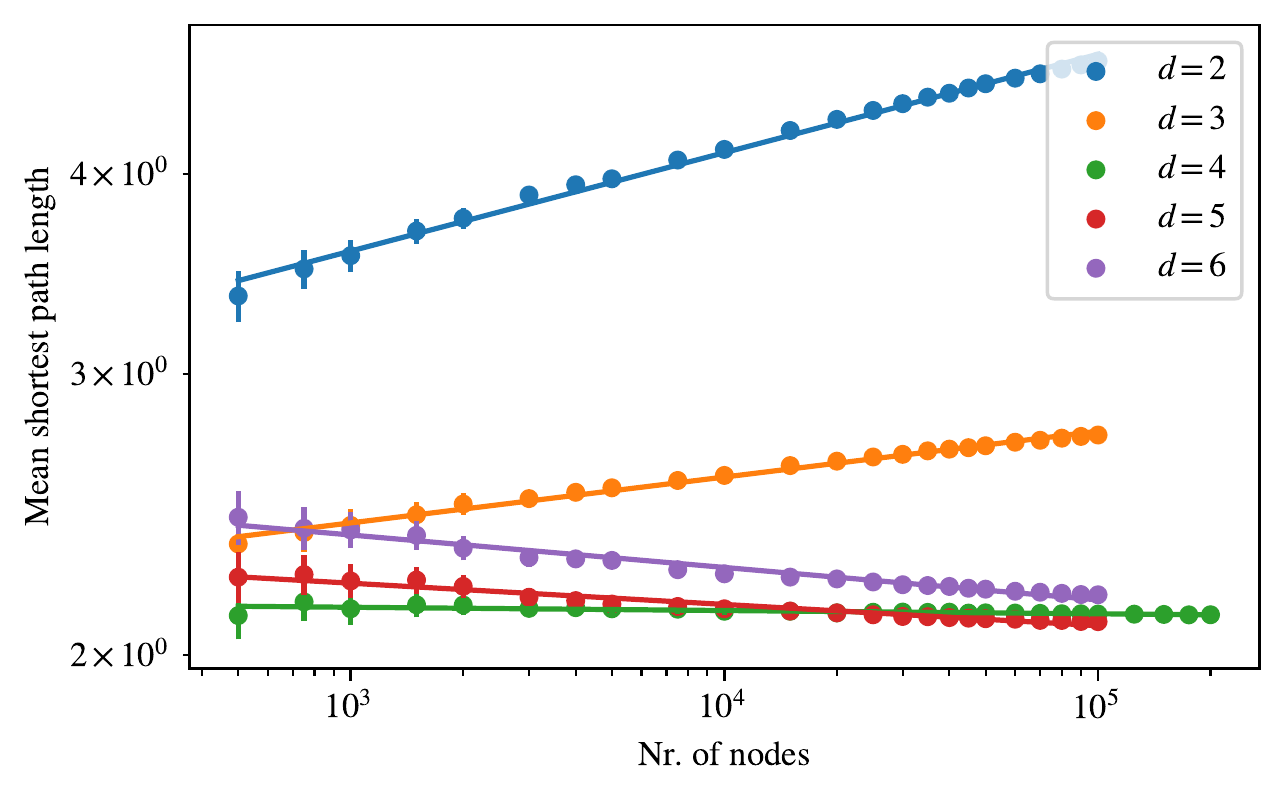}
\caption{\label{fig:causalsetspl}Average shortest path length computed for sprinklings into  $d$ dimensional Minkowski spacetime as a function of the number of nodes. Here $d$ is the number of spacetime dimensions, the spatial dimension accordingly is $d-1$. The average is computed over randomly chosen nodes that comprise 10\% of the causal set. Errorbars indicate the estimated standard error of the mean.}
\end{figure}

We obtain a network determined by the Minkowski metric by borrowing concepts from the causal set approach to quantum gravity \cite{Bombelli:1987aa}, see \cite{Sorkin:2003bx,Dowker:2006wr,Dowker:2013dog,Surya:2019ndm} for reviews. Here, we do not focus on the deep quantum structure of spacetime. Instead, we use concepts from the causal set approach to quantum gravity in a classical context to compare to real-world networks, following up on a first such comparison in \cite{Krioukov:2012mn}.
A causal set is a network that encodes the causal structure of a spacetime with a Lorentzian-signature metric. Its nodes are spacetime points and its (directed) edges stand for causal connections between those points. The causal connections are determined by a Lorentzian metric.\\
Instead of providing the most general definition of a causal set (see \cite{Sorkin:2003bx,Dowker:2006wr,Dowker:2013dog,Surya:2019ndm} for that), we focus on the construction of a causal set from Minkowski spacetime. To obtain the nodes of the network, we use a random sprinkling process: we draw the coordinates for $N$ spacetime points from a uniform distribution. This random distribution of spacetime points preserves Lorentz invariance in a statistical sense\footnote{This statement holds because a random distribution does not select a frame. To generate a sprinkling, the number of points in a volume follows from a Poisson distribution. Here, we generate a representative ensemble member. Lorentz invariance in a statistical sense strictly speaking only holds in the interior of a causal set. For a finite causal set, the shape of the boundary is deformed by a Lorentz boost, thus breaking Lorentz invariance.}.
Given a set of spacetime points in $d$-dimensional Minkowski spacetime, the edges are inferred from the Minkowski metric. To obtain the causal set that encodes the causal structure of Minkowski spacetime, no edges are drawn between points at spacelike separation.
Once all nodes that are causally connected have been linked by an edge, we only consider the irreducible edges, i.e., those edges that cannot be inferred from transitivity. The spacetime-nearest neighbors, i.e., the spacetime points at a fixed timelike distance,  approximate a hyperboloid, in contrast to the sphere in the Euclidean case. Accordingly, a node in a causal set has a high degree\footnote{This statement does not depend on our use of a sprinkling. For instance, if the spacetime points form a regular lattice with non-fine-tuned lattice spacing, there is an infinite number of points at null distance, which correspond to the nearest neighbors of a point.}.
Because the edges connect nearest neighbors at small \emph{spatiotemporal} distance, they stretch over large \emph{spatial} distances, cf.~Fig.~\ref{fig:sprinkling} and Fig.~\ref{fig:sprinklingmetrics}. The network is non-local according to a Euclidean metric.\\
The spectral dimension on causal sets was first evaluated in \cite{Eichhorn:2013ova}, followed by continuum approximations \cite{Carlip:2015mra,Belenchia:2015aia} which extrapolate to below the discreteness scale and do therefore not capture the spectral dimension of a causal set at small diffusion times.
Here, we repeat the analysis in \cite{Eichhorn:2013ova} to provide a template for comparisons with real-world networks.
When setting up the diffusion process, we neglect the directedness of the edges \footnote{Ref.~\cite{Eichhorn:2013ova} introduced a spectral dimension based on a causal diffusion process. It shares the characteristics of the diffusion process we consider here.}. Therefore, the diffusion process does not have a direct interpretation as a physical process on a discrete spacetime. \\
The spectral dimension of a causal set that embeds into $d$-dimensional Minkowski spacetime does not automatically reflect this dimensionality, cf.~Fig.~\ref{fig:diffcsnocutoff}. The large degree, that even grows with the overall spacetime volume, results in a large spectral dimension at small $\sigma$ and a quick subsequent drop, because equilibration sets in quickly.\\
In \cite{Eichhorn:2017djq}, this behavior of the spectral dimension was linked to a conjectured property of quantum gravity, asymptotic silence, following the proposal in \cite{Carlip:2015mra}, see also \cite{Carlip:2009kf,Carlip:2019onx}. \\
The nonlocal structure of a causal set motivates us to ask whether random sprinklings into Minkowski spacetime exhibit a small-world property. The small-world property is typically defined in terms of the average path length as well as the clustering coefficient \cite{WS}. Clusters, i.e., three nodes which are pairwise connected by three edges, cannot occur in the Hasse diagram of a causal set.
Instead, we investigate how the average shortest path between two nodes depends on the size of the network, i.e., the number $N$ of nodes. We find a logarithmic growth for $d=2,3$, cf.~Fig.~\ref{fig:causalsetspl}, characteristic for small-world networks.  For $d=4$ the average path length becomes approximately independent of $N$. At larger $d$, it decreases, cf.~Tab.~\ref{tab:fits}. Adding new nodes hence decreases the distance between two randomly chosen nodes. This result is in qualitative agreement with a simple continuum spacetime approximation, see App.~\ref{app:scaling}. This indicates that for $d \geq 4$, the resulting graphs are more strongly connected than typical small-world graphs\footnote{This might have interesting implications for the navigability of such graphs. See Ref.~\cite{Cunningham:2017afo} for a study of the navigability of curved causal sets.}.
\begin{table}
\centering
\begin{ruledtabular}
\begin{tabular}{lrr}
$d$ & intercept &  slope \\
\hline
2 &     0.850 &  0.062 \\
3 &     0.686 &  0.029 \\
4 &     0.776 & -0.002 \\
5 &     0.889 & -0.013 \\
6 &     1.007 & -0.020 \\
\end{tabular}
\end{ruledtabular}
\caption{\label{tab:fits}
Coefficients of a linear fit of  the relation between $\log N$ and the log of the mean shortest path length.
}
\end{table}
In the sense of networks, our universe (when the small spacetime curvature due to the cosmological constant is neglected) is therefore even more highly connected than a small-world network. This result -- although somewhat remarkable -- does not have direct physical implications, as we achieve it by ignoring the directedness of edges, i.e., disregarding the causal ordering of spacetime events.\\
For the analysis of networks, the small-world type causal sets in two and three dimensions are of more interest, because the relevant dimensionality for the embedding of static real-world networks are two or three spatial dimensions.
\subsection{Hybrid spacetime signature networks, their large but finite degree and their spectral dimension}
\begin{figure}[!t]
\includegraphics[width=\linewidth]{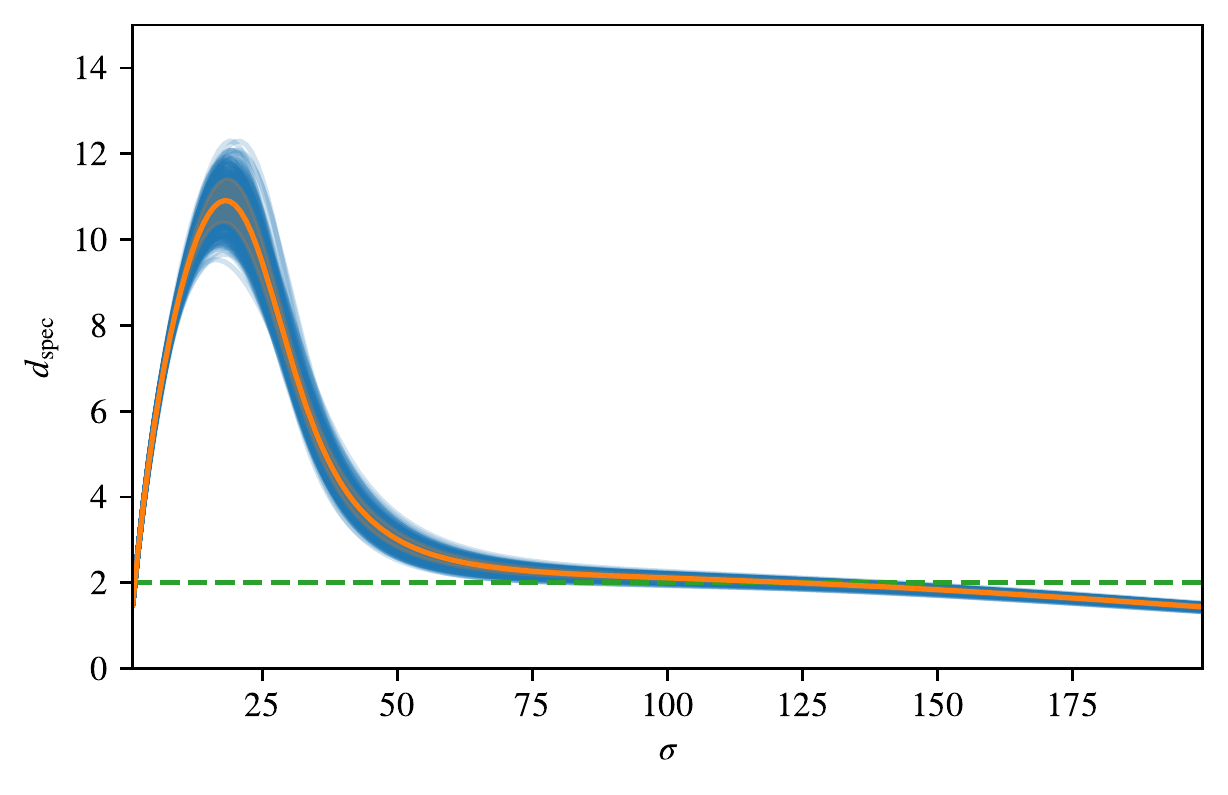}
\caption{\label{fig:causalsetds}Spectral dimension for a causal set that is a sprinkling into 1+1-dimensional Minkowski spacetime with Euclidean cutoff $L=30$ and $N=10^6$ nodes. This corresponds to a  hybrid-signature network, i.e. a network constructed with hybrid spacetime signature.}
\end{figure}

For diffusion on a finite causal set, equilibration sets in very quickly, making it difficult to extract the information on the underlying dimension. In \cite{Eichhorn:2013ova} it was proposed to introduce a large-scale cutoff that prevents a very fast diffusion to the boundary of the network. The cutoff $L$ is imposed according to \emph{Euclidean} distance on a network on which edges are drawn according to \emph{Minkowski} distance. The cutoff removes any edges that span a Euclidean distance larger than $L$. The resulting network is in effect based on a hybrid signature, cf.~Fig.~\ref{fig:sprinklingmetrics}.\\
We now focus on $d=1+1$, as this suffices to exhibit all salient properties, see \cite{Eichhorn:2013ova} for a higher-dimensional case. Due to the large degree, the spectral dimension exhibits a peak at small $\sigma$. At larger $\sigma$, the cutoff results in a plateau in the spectral dimension at $d_{\rm spec} \approx 2$, before equilibration sets in at large $\sigma$, cf.~Fig.~\ref{fig:causalsetds}. \\
In the context of quantum gravity, it is intriguing to observe certain similarities between the phase diagram of a particular subclass of causal sets \cite{Surya:2011du,Glaser:2017sbe,Cunningham:2019rob}, the so-called 2d-orders, and the phase diagram of networks \cite{2002PhRvE..66c7102S} which are closely related to the hybrid-spacetime signature networks that we explore. Both phase diagrams feature a regular and a random phase. The hybrid-signature case additionally features an intermediate small-world phase and the phase transitions are of higher order.
\subsection{Watts-Strogatz model}
\begin{figure*}
\includegraphics[width=\linewidth]{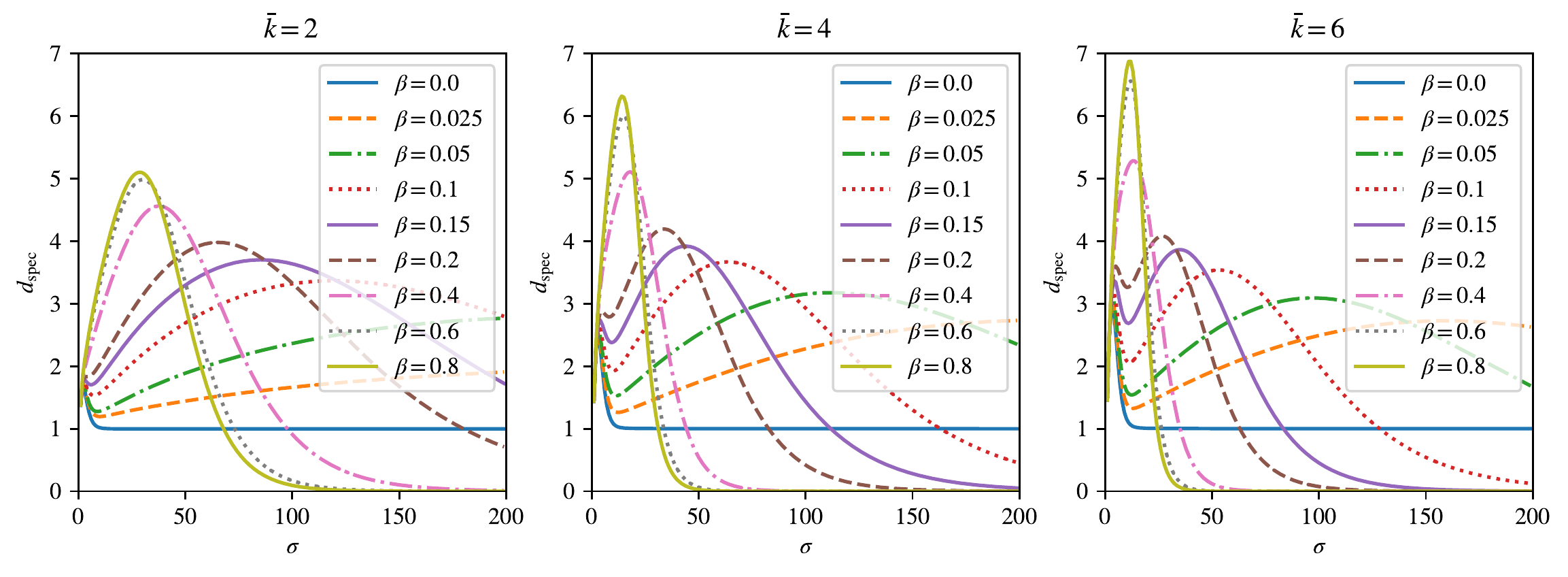}
\caption{\label{fig:WS}
Spectral dimension for a Watts-Strogatz graph with $N=1000$ nodes with varying values for the average degree $\bar{k}$
and the rewiring parameter $\beta$. For each of these type of graphs we are averaging over five randomly generated graphs and 25 start nodes per graph.}
\end{figure*}
\begin{figure}[!t]
\includegraphics[width=\linewidth]{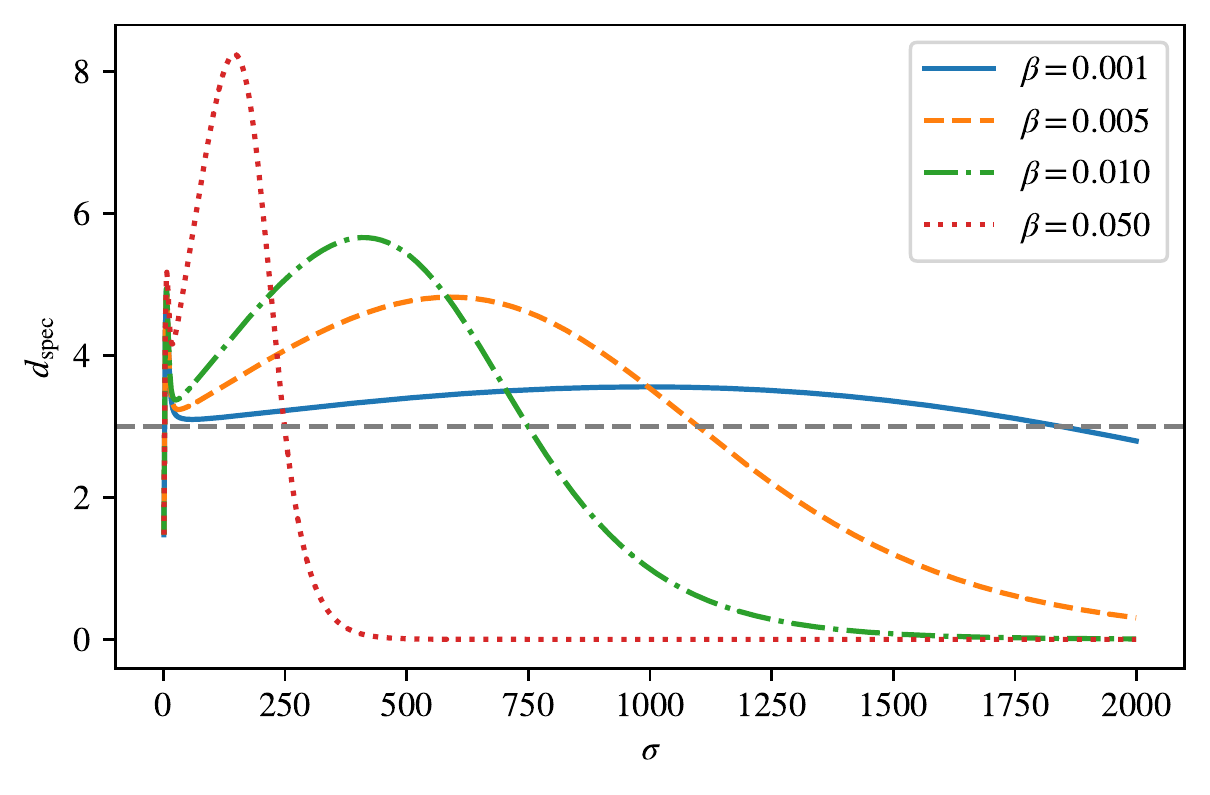}
\caption{\label{fig:WShigherd_var_beta}Mean spectral dimension for a $3d$ generalization of a Watts-Strogatz graph. Each node is connected to all neighbors within a Euclidean distance $\leq 2$. Edges are then rewired with the rewiring probability $\beta$. The gray dashed line marks $d_\text{spec}=3$.}
\end{figure}

The Watts-Strogatz model \cite{WS} starts from a one-dimensional ring as in Sec.~\ref{sec:grid}. Subsequently, the graph-local neighborhood is changed to include spacetime-non-local connections. This is implemented by rewiring each node in the original ring with probability $\beta$, such that the number of edges stays constant as the network becomes more and more small-world. The effect can be seen in the left panel of Fig.~\ref{fig:WS}, where the spectral dimension exhibits a similar scale dependence to the case of the ring with the added edges in the lower panel of Fig.~\ref{fig:ring}.
The central and right panel of Fig.~\ref{fig:ring} show the Watts-Strogatz model starting from a ring with not just nearest neighbors, but also next-nearest and next-next-nearest neighbors. In these cases and for an appropriate choice of $\beta$, the spectral dimension shows two maxima, one at small $\sigma$ and another at intermediate $\sigma$.  We conjecture that the two maxima occur because the graph exhibits two distinct length scales associated to the onset of non-locality: first, the neighborhood of each node becomes increasingly non-local, as $\beta$ increases. This results in the first peak in the spectral dimension. The second maximum is associated to  efficient connections that are highly nonlocal, i.e., link two nodes which would be very far from each other without the extra connection. At low $\beta$, only few of these connections exist, and on average, the diffusion process requires a large number of steps to reach these efficient connections. Once they are reached, the number of nodes that can be reached in the next few steps of the diffusion process is significantly larger. Therefore, the spectral dimension peaks at a second maximum at these larger diffusion times.\\
For larger $\beta$, there are many efficient, non-local connections. These result in a large number of nodes that can be reached in the first few steps of the diffusion process. At the same time, they speed up the onset of equilibration. Accordingly, the spectral dimension peaks at high values at low diffusion times and then plummets to zero.
We generalize the transition between a ring and a Watts-Strogatz graph to higher dimensions, see also \cite{2002PhRvE..66c7102S} for a similar construction.
We start out with nodes on a regular lattice embedded in $\mathbb{R}^d$. We connect each of the nodes to all nodes at a Euclidean distance smaller than a cutoff.  Subsequently, each of the edges is rewired to connect to a random other node with probability $\beta$. As $\beta$ increases, the network starts to exhibit small-world properties.\\
Just like in the one-dimensional case, i.e., the original Watts-Strogatz model, the spectral dimension can feature two maxima: for an appropriate range of $\beta$, there is a sharp peak in $d_{\rm spec}$ at small $\sigma$, followed by a sharp drop and a relatively slow increase towards a second maximum at larger $\sigma$. Which of the two maxima is the global one depends on the choice of $\beta$, cf.~Fig.~\ref{fig:WShigherd_var_beta}. \\
At sufficiently large $\beta$, the spectral dimension does not exhibit plateau-like behavior. Nevertheless, the regime with two distinct maxima appears to feature a height of the maxima that increases with the original embedding dimension.
The generalization of the Watts-Strogatz model to higher dimensions illustrates, how even in the presence of moderate small-world behavior, the underlying dimensionality remains relevant.
The model provides another example of a network that encodes the transition from a local network based on Euclidean distance to a non-local network and exhibits the imprint of nonlocality on the spectral dimension.
\section{Scale-dependent spectral dimension for real-world networks}\label{Sec:realnetworks}
All real-world networks that exist embedded in the physical world (e.g., road networks, the fibre optics cables connecting servers and end-users on the internet, the synapses in a brain) are embedded in a space of given dimensionality.
We aim to test how strongly such spatial real-world networks are constrained by the geometric and topological properties of the embedding space. In particular, for networks which are embedded in two or three dimensions, such as, e.g., a road-network, or the neural network of a biological brain, we ask: does the embedding influence the geometric properties of the network itself? In particular, is the embedding dimension encoded in the connectivity of the network and thus the spectral dimension?\\
We use the spectral dimension on the  synthetic networks from the previous section as templates.
We will find that the Euclidean-signature spectral dimension provides a good template for road networks, which do not exhibit small-world properties.
We will find that the Watts-Strogatz model template plays a role for networks that exhibit small-world properties (in the sense of small network diameter), like networks encoding the internet architecture and the neural networks of biological brains. Finally, we will find that those networks also show spectral dimensions compatible with the Lorentzian as well as the hybrid template.
This suggests an -- to our best knowledge so far only explored in \cite{Krioukov:2012mn} -- connection between the structures underlying Lorentzian discrete quantum gravity and certain real-world networks.
\subsection{Road networks}
\begin{figure}[!t]
\includegraphics[width=\linewidth]{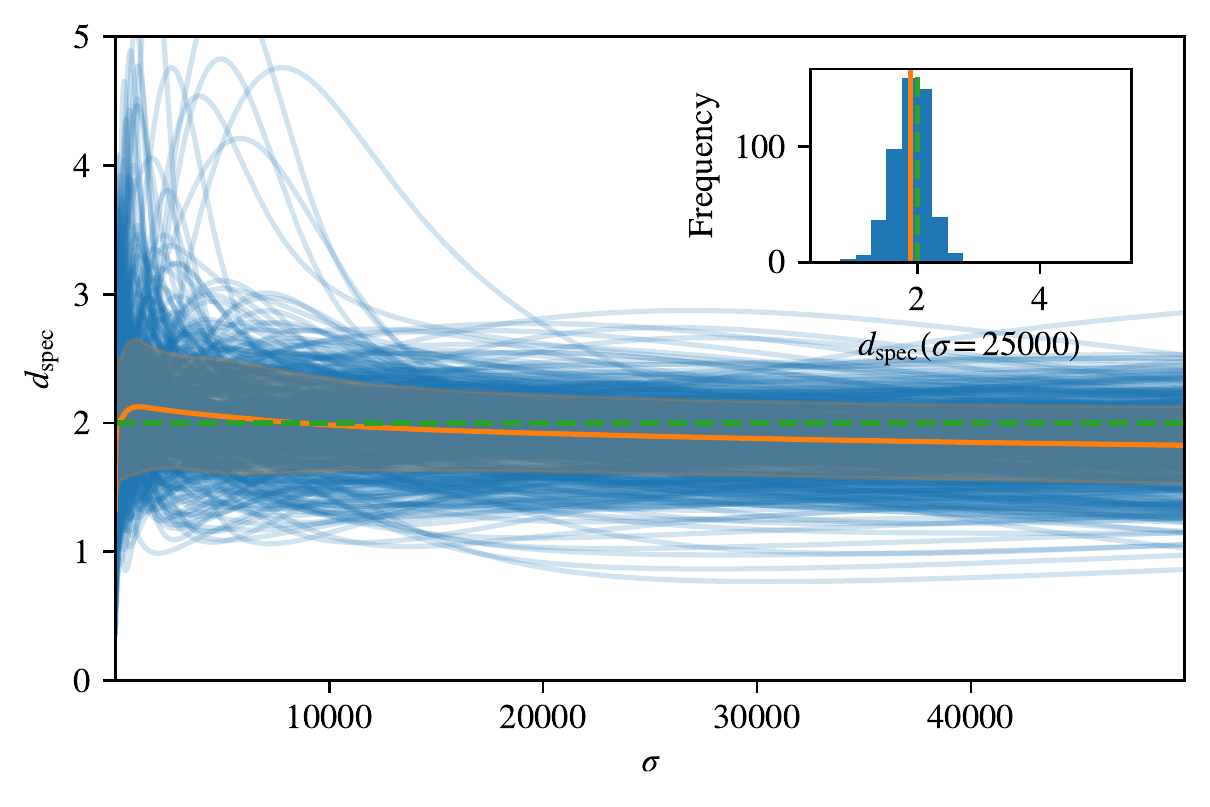}
\caption{\label{fig:streetsofpenn} Spectral dimension for a network representing the road network of Pennsylvania. The blue lines correspond to the spectral dimension for different start nodes, the orange line represents the average over start nodes, the orange band represents the estimate of the standard deviation. The inset is a histogram showing the distribution over start nodes for a fiducial $\sigma$. The spectral dimension plateaus at $d_\text{spec} \approx 2$ (green dashed line).}
\end{figure}

\begin{figure}[!t]
\includegraphics[width=\linewidth]{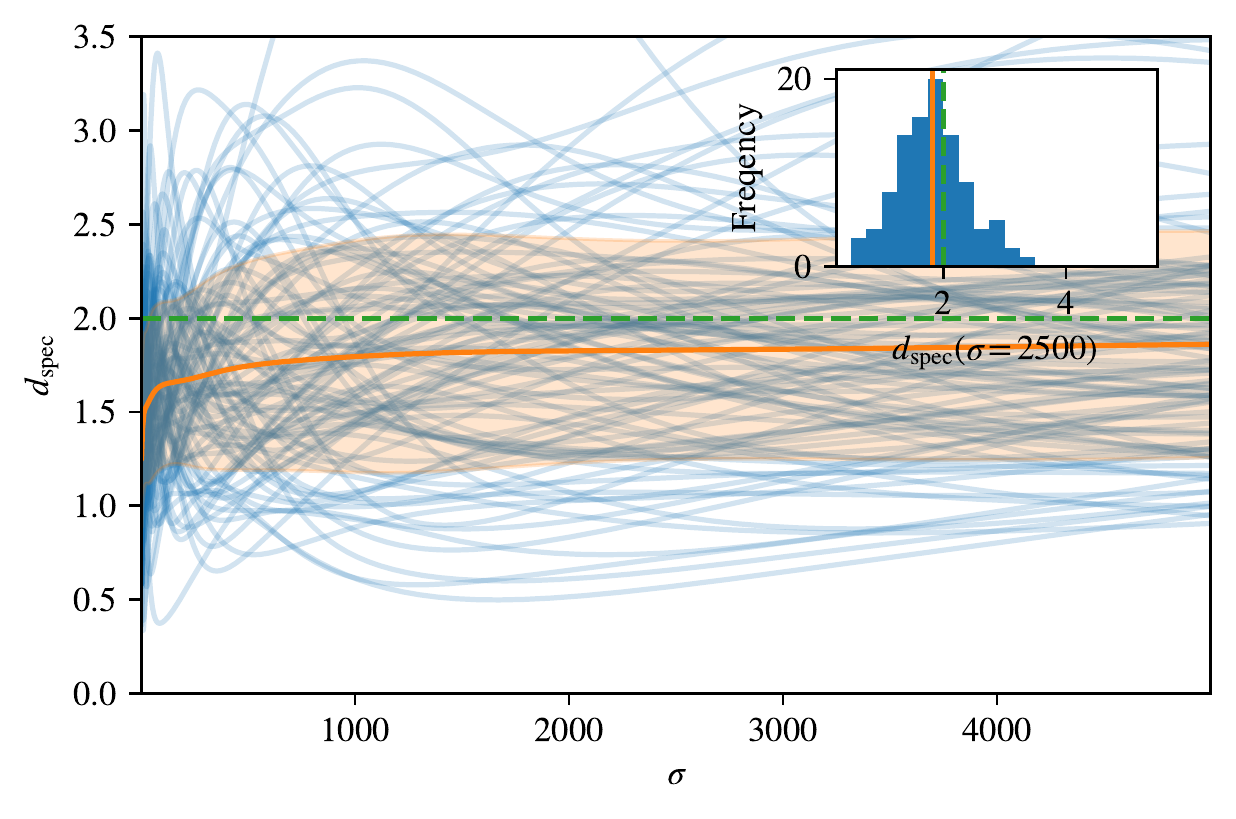}
\caption{\label{fig:streetsofeurope} Spectral dimension for a network representing the road network of Europe. The spectral dimension again plateaus at $d_\text{spec} \approx 2$.
}
\end{figure}

Road networks are a paradigmatic example of a real-world network embedded in a space. This space is (approximately) two-dimensional. Here, we test whether the spectral-dimension templates with Euclidean, Lorentzian or hybrid signature match the spectral dimensions of road networks.\\
Explicit examples are given by the network of streets in Pennsylvannia, cf.~Fig.~\ref{fig:streetsofpenn} and the network of  major roads in Europe, cf.~Fig.~\ref{fig:streetsofeurope}. The nodes of these networks correspond to intersections, the links to roads.
Both networks are inhomogeneous in that they do not have a constant number of nearest neighbors across the network. The inhomogeneity causes the spectral dimension to depend on the starting point of the random walk, cf.~blue lines in Fig.~\ref{fig:streetsofpenn} and Fig.~\ref{fig:streetsofeurope}. Even at large $\sigma$, a given individual random walker can measure a spectral dimension that can vary between 1 and 3. Despite the large variations at the level of individual random walks, the dimensionality of the embedding space is captured quite well at the average level:
we average over the starting points, resulting in $d_{\rm spec} \approx 2$ over the entire range of $\sigma$ we consider here.
For the road network of Pennsylvania, we simulate random walks with up to $5 \cdot 10^4$ steps, for the road network of Europe, we simulate $5\cdot 10^3$ steps. \\
The result for the spectral dimension agrees with the topological dimension of the embedding space, highlighting how the embedding of road networks in two dimensional spaces constrains the morphology of the network.\\
The spectral dimension is well-captured by the template provided by a network obeying Euclidean distances, as seen by comparing Fig.~\ref{fig:streetsofpenn} and Fig.~\ref{fig:streetsofeurope} to Fig.~\ref{fig:ring}. The Minkowski and the hybrid template are not relevant to describe the spectral dimension for road networks. We tentatively conclude that to a good approximation road networks connect nearest neighbors in a Euclidean metric and do not exhibit small-world properties.
\subsection{Overcoming geometric constraints: Internet architecture and brain networks}
Road networks do not constitute small-world networks. Accordingly, information propagates slowly and the distance between two random nodes grows faster than logarithmically with system size. Therefore, such networks architectures are not suitable for systems which rely on an efficient propagation of information. Instead, the physical architecture of the internet \cite{1999Natur.401..130A,bu2002distinguishing} as well as various brains \cite{Hilgetag,Sporns,Bassett,Salvador,Bassett2,Bassett3} exhibit small-world properties. Yet, these networks are also embedded into three- and two-dimensional space, respectively. The restrictions arising from the geometric embedding (which can result in costliness of long-range connections) have to be balanced against the need for high connectivity.
Here, we investigate the spectral dimensions of these networks to find out whether they match or resemble that of the networks embedded in a spacetime with Lorentzian or hybrid metric.\\
Due to the inhomogeneous nature of the underlying networks, different nodes differ in their local neighborhood. Accordingly, we do not expect to find the same result for the spectral dimension for random walks starting at different nodes. In the following, we distinguish classes of starting nodes, inspired by the various templates discussed in the previous section. We search for the presence of four distinct classes:
\begin{enumerate}
\item Walks in the \emph{non-local class} exhibit an immediate rise in the spectral dimension, and then quickly decay towards $d_\text{spec} \approx 0$. Such a spectral dimension can be obtained for geometric networks constructed with the Minkowski metric, or for non-geometric networks that start with a Euclidean network but contain many non-local edges in addition, cf.~lower panel Fig.~\ref{fig:ring}, Fig.~\ref{fig:diffcsnocutoff}. It can also be obtained for a Watts-Strogatz model with high $\beta$, cf.~Fig.~\ref{fig:WS}. This class signals a high connectivity of the network. Therefore we expect this class to occur in small-world networks.
\item Walks in the \emph{hybrid class} exhibit an immediate rise, but then a fall towards an intermediate plateau at finite $d_\text{spec} > 0$. Such a spectral dimension can be obtained for geometric networks constructed with a hybrid metric, cf.~Fig.~\ref{fig:causalsetds}. This class signals a highly connected region embedded into a larger less connected structure.
\item Walks in the \emph{local class} exhibit a low spectral dimension at low $\sigma$, followed by a rise towards a plateau at intermediate $\sigma$. Such a spectral dimension can be obtained for geometric networks with Euclidean signature, cf.~upper panel in Fig.~\ref{fig:ring}.
\item Walks in the \emph{Watts-Strogatz class} exhibit two maxima in their spectral dimension. This is reminiscent of the Watts-Strogatz graph with relatively low $\beta$. This class signals the presence of a few long-range ``shortcuts'', cf.~Fig.~\ref{fig:WS}.
\end{enumerate}
The first three classes are geometric ones, in which the embedding geometry imprints on the network, and in turn information on the embedding dimensionality can be extracted from the random walks in the second and third class.\\
Notably, we will not find a significant number of walks exhibiting a spectral dimension in the local class. This indicates that there is no significant fraction of nodes with an approximately Euclidean neighborhood. Instead, both other geometric classes as well as the Watts-Strogatz class appear in real-world networks with small-world property, as we will show now.
\subsubsection{Internet architecture}
\begin{figure*}[!t]
\includegraphics[width=0.5\linewidth]{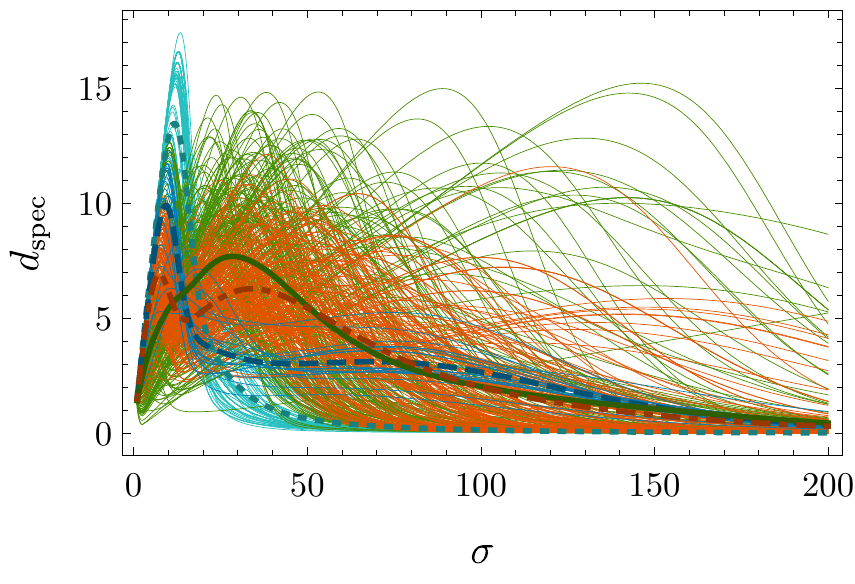}\\
\includegraphics[width=0.45\linewidth]{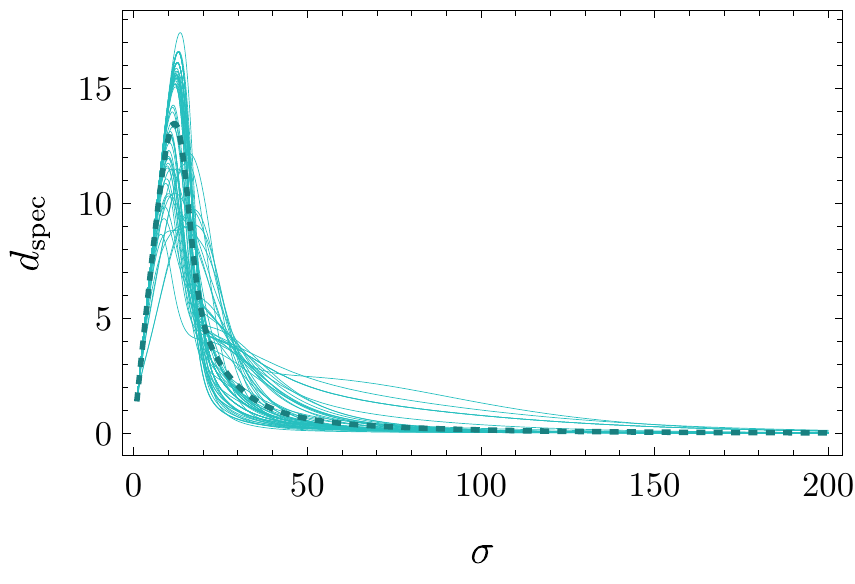}\quad
\includegraphics[width=0.45\linewidth]{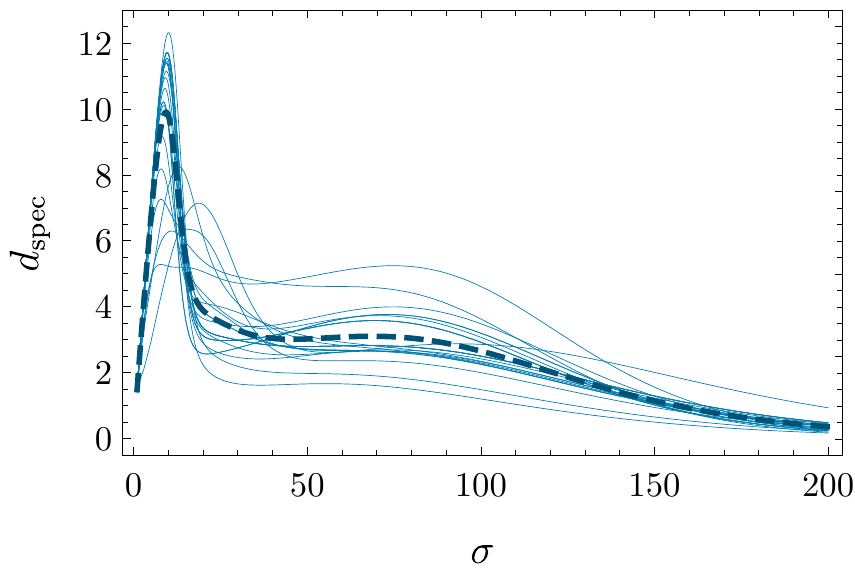}\\\
\includegraphics[width=0.45\linewidth]{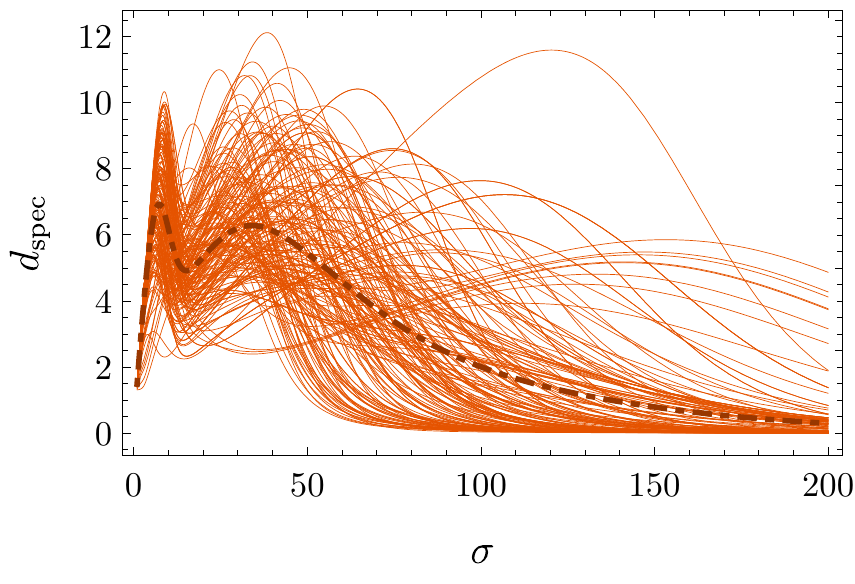}\quad
\includegraphics[width=0.45\linewidth]{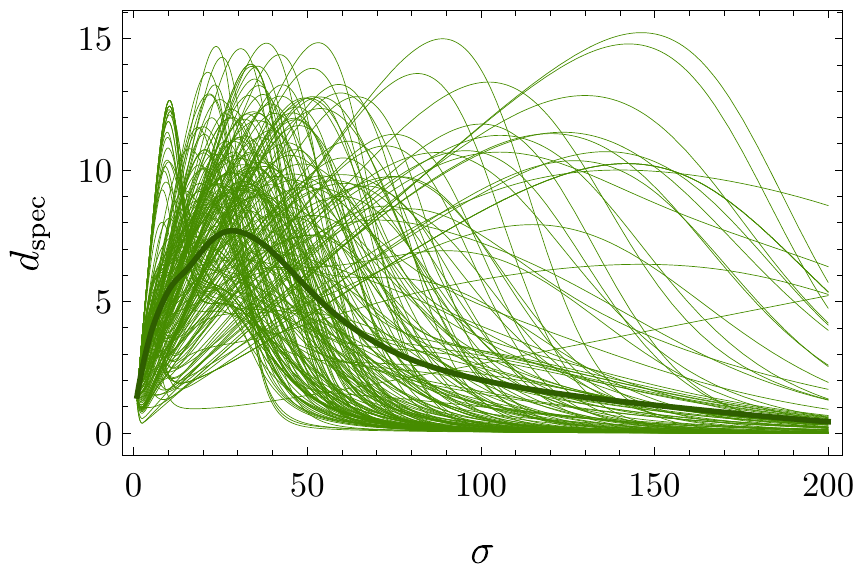}
\caption{\label{fig:internet}Spectral dimension for a graph that describes the structure of the internet according to the  CAIDA database. The upper panel shows all starting points and the average spectral dimension for four classes of diffusers: non-local (cyan, shown in the left panel in the second row, average in dotted),
hybrid (blue, shown in the right panel in the second row, average in blue dashed),
Watts-Strogatz-like (shown in the left panel in the lower row, average in red dot-dashed),
and all remaining diffusers that peak at larger $\sigma$ (green, shown in the right panel in the lower row, average in continuous).}
\end{figure*}
The internet does not correspond to a unique network. In fact, different networks can be associated with it, depending on the definition of node and link. For instance, the information infrastructure of the internet is associated with a network in which webpages are nodes, and hyperlinks are edges. Alternatively, the physical architecture of the internet can be described as a  network in which so-called autonomous systems -- i.e., individual administrative networks -- are nodes and physical data connections are edges. This network is  expected to be constrained by its embedding into a space. We aim at finding out how strongly the embedding is imprinted on the spectral dimension. We expect a clear difference to the spectral dimension of a regular grid, because efficient connectivity is an important goal for the physical internet architecture. It can be achieved by laying short-cuts across larger spatial distances.
The CAIDA database provides snapshots of the autonomous systems and their connections comprising (parts of) the internet.
The network constructed from  such a snapshot is rather inhomogeneous, cf.~Fig.~\ref{fig:internet} and Tab.~\ref{tab:classes_count}. We distinguish four classes of walks, three of them discussed above:
\\
First, we find a large number of walks in the non-local class. Given the high degree of connectivity of the internet this is expected. \\
Second,  a fraction of the walks fall into the hybrid class with a plateau in the spectral dimension at $d_{\rm spec} \approx 3$. To the best of our knowledge, this is the first example of a real-world network for which a hybrid-signature spacetime is relevant.
It is perhaps not so surprising that the network we investigate retains an imprint of the embedding space. However, one might have expected that a Euclidean metric is the relevant one in this context. We do not find walks that resemble the walk in a Euclidean metric, i.e., the local class is not present in this network.\\
Third,  a significant fraction of walks fall into the Watts-Strogatz class. Among the templates that we have constructed that start from a grid in $d$ dimensions, the three-dimensional version, cf.~Fig.~\ref{fig:WShigherd_var_beta}, appears to resemble the result closest in terms of the height of the two maxima.  While this is consistent with our findings for the second class of walks, we have not investigated quantitatively whether there are degeneracies in the spectral dimension when the average degree, embedding dimension and rewiring probability are varied. \\
Fourth, there is a remainder class showing a spectral dimension resembling that of Watts-Strogatz models for $\beta$ smaller than the previous class, where the spectral dimension increases towards a peak at larger $\sigma_{\rm max}$. As is evident from the lower right panel in Fig.~\ref{fig:internet}, there is a significant spread in $\sigma_{\rm max}$. Accordingly, it might be possible to decompose this class further into meaningful subclasses.\\
Overall, we conclude that the network based on the internet architecture carries imprints of its embedding into space. The geometric class features a plateau in $d_{\rm spec} \approx 3$, with a large uncertainty of the value. The relevant geometric network is a hybrid-signature one. It appears to capture the appropriate degree of non-locality to provide a spectral dimension which is very similar to that of the internet architecture.
\begin{table}
\centering
\begin{ruledtabular}
\begin{tabular}{lcccc}
Graph & Class 1 & Class 2 & Class 4 & Remainder \\
\hline
CAIDA & 73 & 24 & 194 & 209 \\
Brain of Drosophila & 1274 & 343 & 108 & 275 \\
Brain of Mouse & -- & 30 & -- & 70 \\
\end{tabular}
\end{ruledtabular}
\caption{\label{tab:classes_count}
Number of walks in each of the four classes for the various graphs that we consider. For a definition of the classes see the main text.
}
\end{table}
\begin{figure}[!t]
\includegraphics[width=\linewidth]{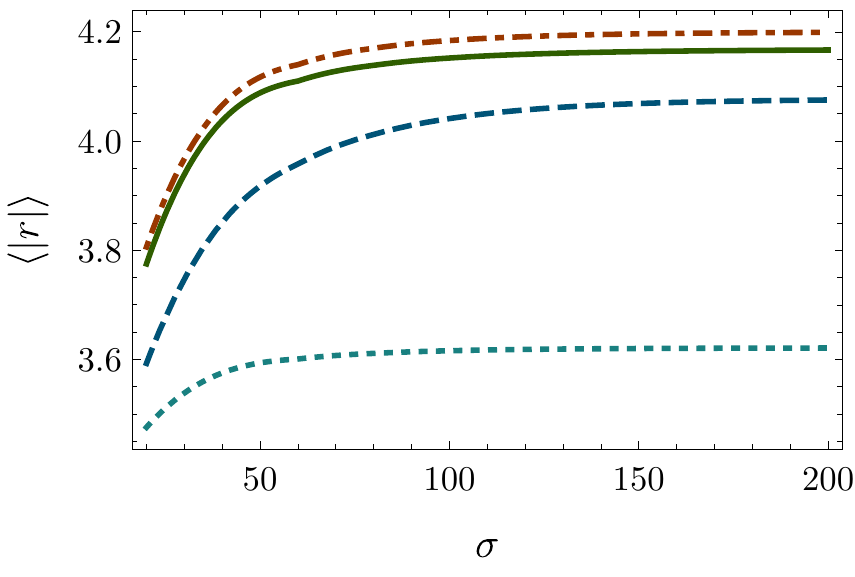}
\caption{\label{fig:internetr}The expectation value of the absolute value of the network distance (i.e., shortest path between two nodes) to the start node  for a graph describing the structure of the internet. The same classes of diffusers as in Fig.~\ref{fig:internet} are shown: hybrid (blue, average in blue dashed), Watts-Strogatz-like (orange, average in red dot-dashed), non-local (cyan, average in dotted) and all remaining diffusers (green) that peak at larger $\sigma$. There is a marked gap in the asymptotic average distance for the small-world diffusers.}
\end{figure}
The same three classes and the remainder class can also be distinguished in the average distance $\langle |r|\rangle$ of the random walk to the starting point, i.e., the shortest path between the starting node and the node occupied at time $\sigma$, cf.~Fig.~\ref{fig:internetr}. We observe three key points:\\
First, the average distance reaches asymptotic values between 3.5 and 4.2, as expected for a network with $N=33304$ nodes of small-world type, i.e., logarithmic dependence of the average distance on $N$. \\Second, the non-local class asymptotes at the lowest $\langle |r|\rangle$ and reaches that asymptotic value fastest, as expected for a highly non-local network on which equilibration occurs fastest. \\
Third, walks in the Watts-Strogatz class and the  remaining walks -- potentially described by a Watts-Strogatz model at lower $\beta$ -- show $\langle |r|\rangle(\sigma)$ well compatible with each other, strengthening the hypothesis that both components might be captured by Watts-Strogatz-type networks.
\subsubsection{Neural networks}
\begin{figure*}[!t]
\includegraphics[width=0.6\linewidth]{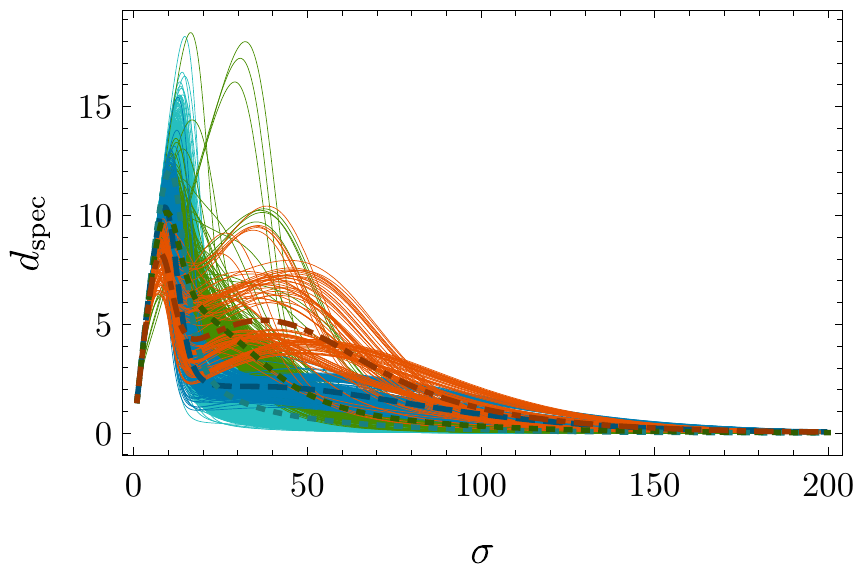}\\
\includegraphics[width=0.45\linewidth]{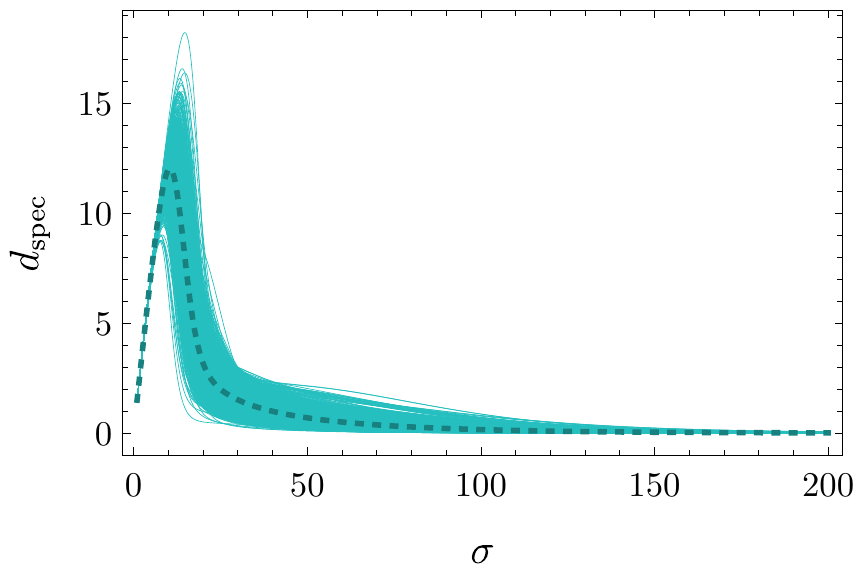}\quad
\includegraphics[width=0.45\linewidth]{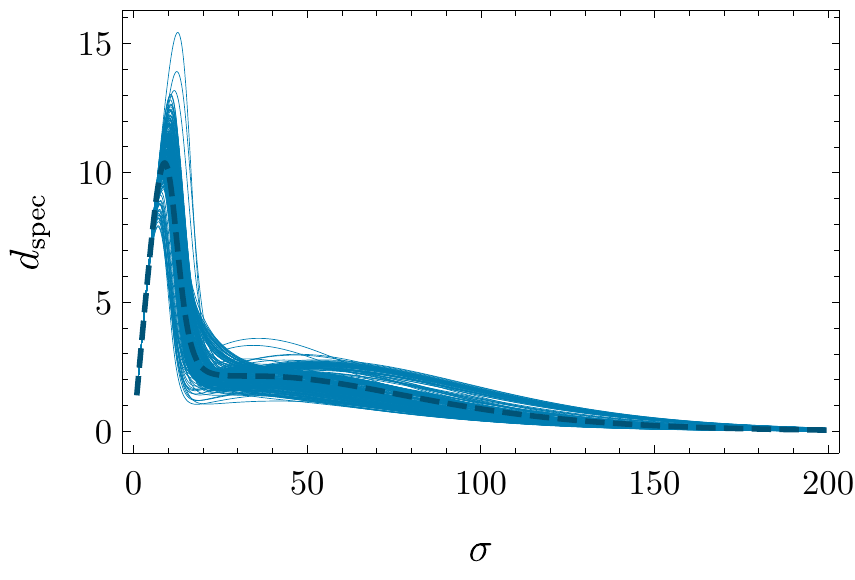}\\\
\includegraphics[width=0.45\linewidth]{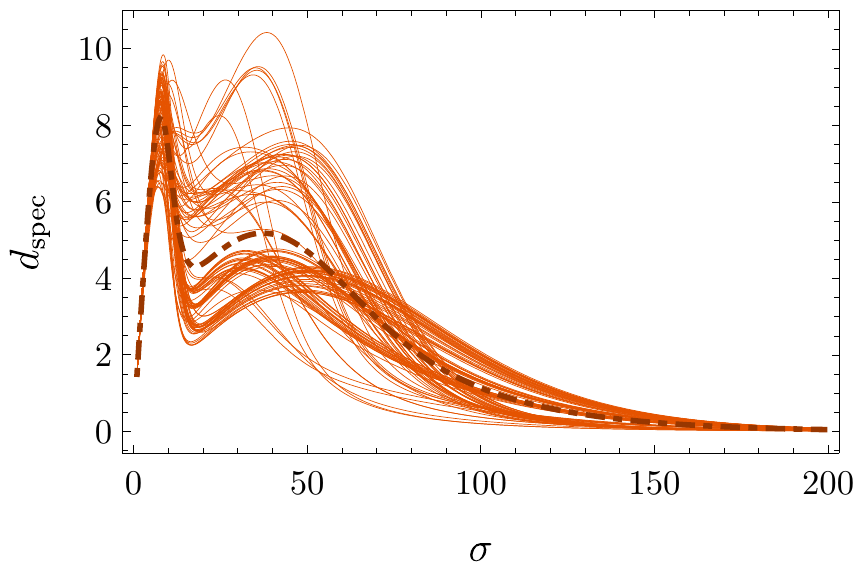}\quad
\includegraphics[width=0.45\linewidth]{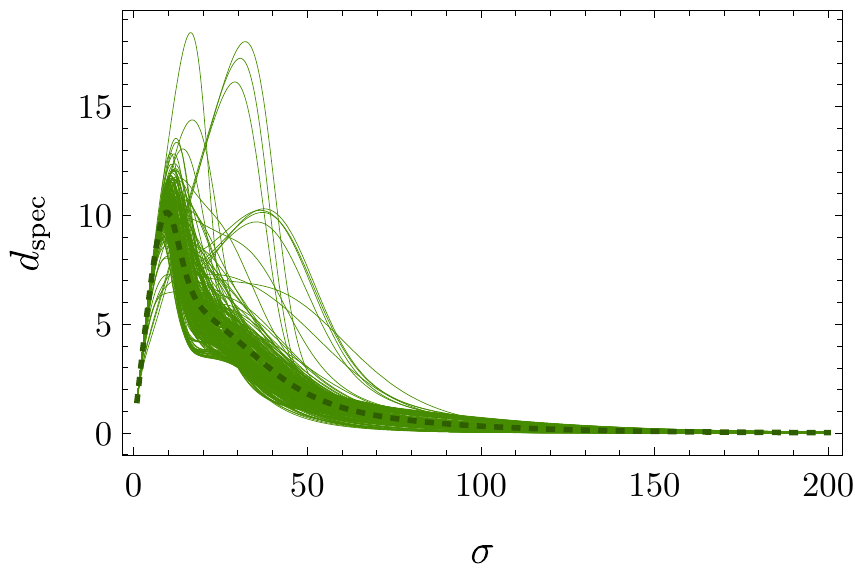}
\caption{\label{fig:drosophila} We show the spectral dimension  for a graph that represents the neural network of a drosophila fly.
Each continuous line is the spectral dimension associated to one of  2000 randomly selected starting nodes.
The upper panel shows all starting points and the average spectral dimension for four classes of diffusers: non-local (cyan, shown in the left panel in the second row, average in dotted), hybrid (blue, shown in the right panel in the second row, average in blue dashed),
Watts-Strogatz-like (shown in the left panel in the lower row, average in red dot-dashed),
and all remaining diffusers that peak at larger $\sigma$ (green, shown in the right panel in the lower row, average in continuous).}
\end{figure*}

\begin{figure}[!t]
\includegraphics[width=\linewidth]{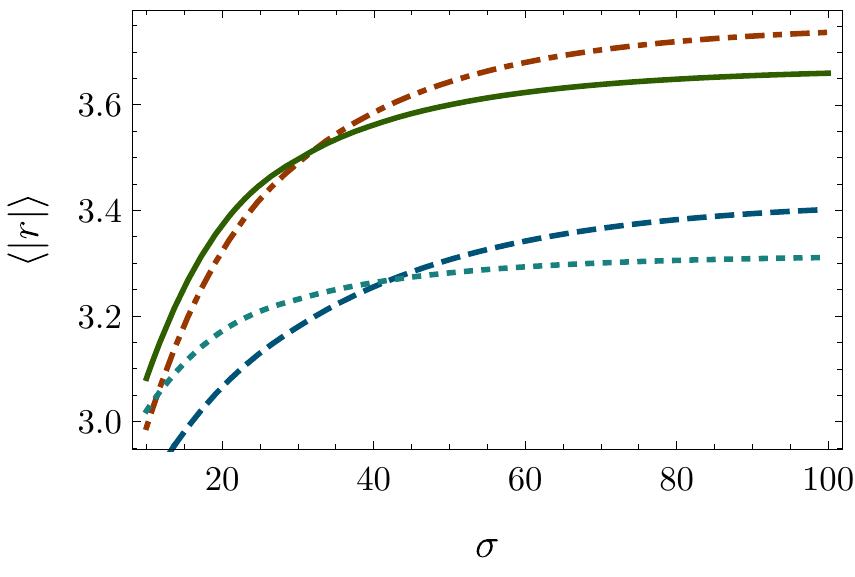}
\caption{\label{fig:drosophilar}The expectation value of the absolute value of the network distance (i.e., shortest path between two nodes) to the start node for a graph that represents the neural network of a drosophila fly. The same classes of diffusers as in Fig.~\ref{fig:drosophila} are shown: hybrid (blue, average in blue dashed), Watts-Strogatz-like (orange, average in red dot-dashed), non-local (cyan, average in dotted) and all remaining diffusers (green) that peak at larger $\sigma$.}
\end{figure}

\begin{figure}[!t]
\includegraphics[width=\linewidth]{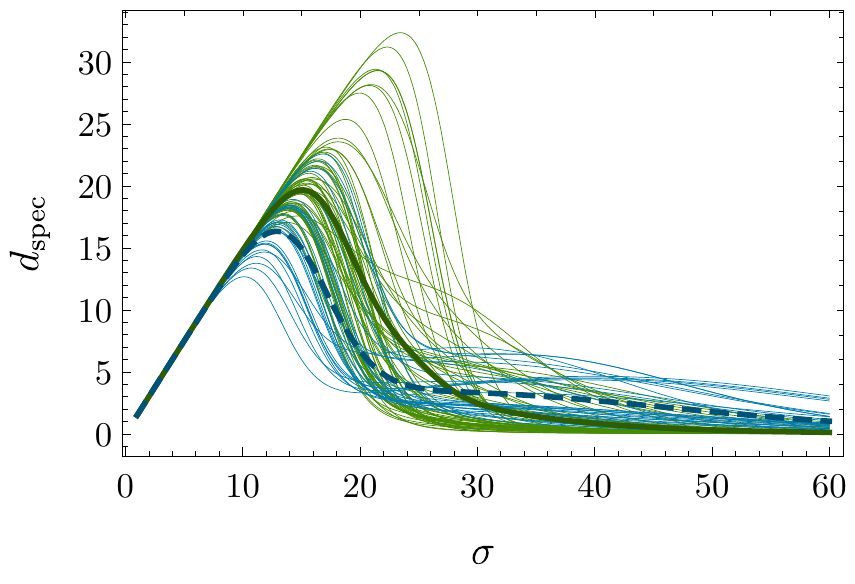}
\includegraphics[width=\linewidth]{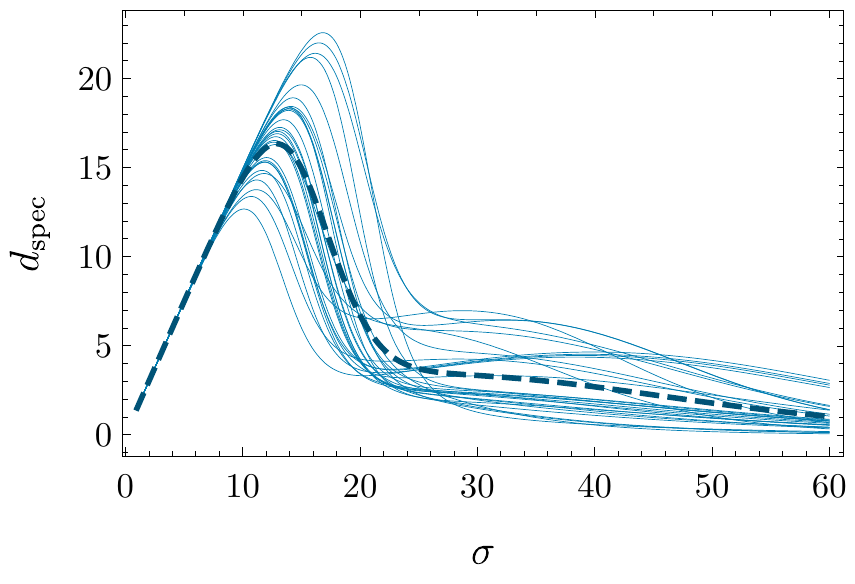}
\caption{\label{fig:rat_voxel_brain} We show the spectral dimension for a graph that represents the neural network of a mouse in a coarse-grained way. We execute a random weighted walk, where the probability depends on the weights of the links. In the upper panel, we show the two classes, first the  nonlocal class that exhibits an early maximum and a subsequent decay (green, average in dark green), second the hybrid class (blue, average in blue, dashed). In the lower panel, we only show the  hybrid class.}
\end{figure}

We consider random walks on the neural network (i.e., the connectome, taken from \cite{xu2020connectome}) of the drosophila fly. \\
Our first nontrivial result is that the three of the four classes of diffusors we identify are the same as those for the CAIDA network, cf.~Fig.~\ref{fig:drosophila}. These three are part of the classes we search for, whereas the local class, based on Euclidean signature, is again absent.
The occurrence of these same three classes in the internet architecture and biological neural networks might signal a deeper similarity between the two networks.\\
Specifically, a large number of walks again belongs to the nonlocal class, with the spectral dimension exhibiting a high initial peak and quick subsequent decay. \\
For the hybrid class of walks the spectral dimension plateaus at $d_{\rm spec} \approx 2$, with a significant spread around this average value. \\
In this case a smaller fraction of walks falls into the Watts-Strogatz class.\\
Finally, the fourth class of walks exhibits a spectral dimension with a single maximum that occurs at larger diffusion times.\\
We conclude that the network carries imprints of its embedding into space. Again, as in the case of the CAIDA network, the hybrid signature network provides the best template to match the hybrid class. In particular, there are no diffusors that behave as in the local class based on Euclidean-signature networks.
The different classes feature asymptotic values for $\langle |r|\rangle$ between 3.2 and 3.8, as is to be expected for a small-world network with $N = 21739$ nodes.  Again, the non-local class shows the fastest approach to the asymptotic value and exhibits the lowest asymptotic value of all four classes, cf.~Fig.~\ref{fig:drosophilar}.
Next, we consider a coarse-grained neural network.
Interestingly, similar classes of random walks are found for this rather different representation of a neural network. Specifically, we consider a mouse brain that is divided into voxels (three dimensional volumes) that represent  the nodes of the network. To obtain the edges of the network,
the correlation between voxels is measured.  If the correlation is larger than a cut-off, then two voxels are linked. The strength of the correlation provides each link with a weight. We implement a weighted random walk that accounts for the weights in the probability to hop along an edge to the next node.
This network encodes the actual neural network at a rather coarse-grained level. Two classes of random walks appear to be robust under the coarse graining from a connectome-network to the voxel-network (we assume that the difference between drosophila and a mouse is irrelevant in this context), namely the hybrid class and the nonlocal class, cf.~Fig.~\ref{fig:rat_voxel_brain}. In contrast, no random walks of the Watts-Strogatz class can be detected in the present case. For the hybrid class, the plateau in the spectral dimension lies at $d_{\rm spec} \approx 3$.\\
Again, as in the other networks in this section, hybrid-signature networks provide a relevant template for this neural network. In contrast, Euclidean-signature networks do not play any role in the diffusion processes on these networks.
In summary, we tentatively conclude that the embedding geometry is imprinted in neural networks.
The first class of walks on the neural networks we considered exhibit the sharp rise and subsequent fast drop of the spectral dimension that is associated with highly connected networks. This class resembles the spectral dimension in Lorentzian-signature networks, but also occurs in any sufficiently non-locally connected network. Therefore, we do not consider this class strong evidence that Lorentzian-signature networks might provide relevant templates for real-world networks. In contrast,  the second class of walks exhibits a spectral dimension characteristic for hybrid-signature networks, where the Minkowski metric provides the relevant notion of distance and nearest neighbors below a Euclidean cutoff. The resulting characteristic curve for the spectral dimension was first observed in the study of a regularized causal-set approach to quantum gravity \cite{Eichhorn:2013ova} and here we discover a ``sprinkling" of such curves in the distinct classes of random walks on networks representing the internet architecture as well as two distinct neural networks. This hybrid-signature curve is determined both by the high degree characteristic of small-world networks and by the dimensionality of the embedding space. In the three cases we consider, the diffusers ``measure" a dimensionality between 2 and 3, which is surprisingly close to the value we expect based on the dimensionality of the embedding spaces (2 for the  internet architecture and 3 for the neural networks).
We conclude that first, such networks are shaped by the embedding space and encode its dimensionality; and second that Euclidean-signature networks do not provide templates for the corresponding spectral dimension. In contrast, Lorentzian-signature and in particular hybrid-signature networks provide relevant templates.
\subsection{Outlook: Spectral dimension of growing networks}
\begin{figure}
\includegraphics[width=\linewidth]{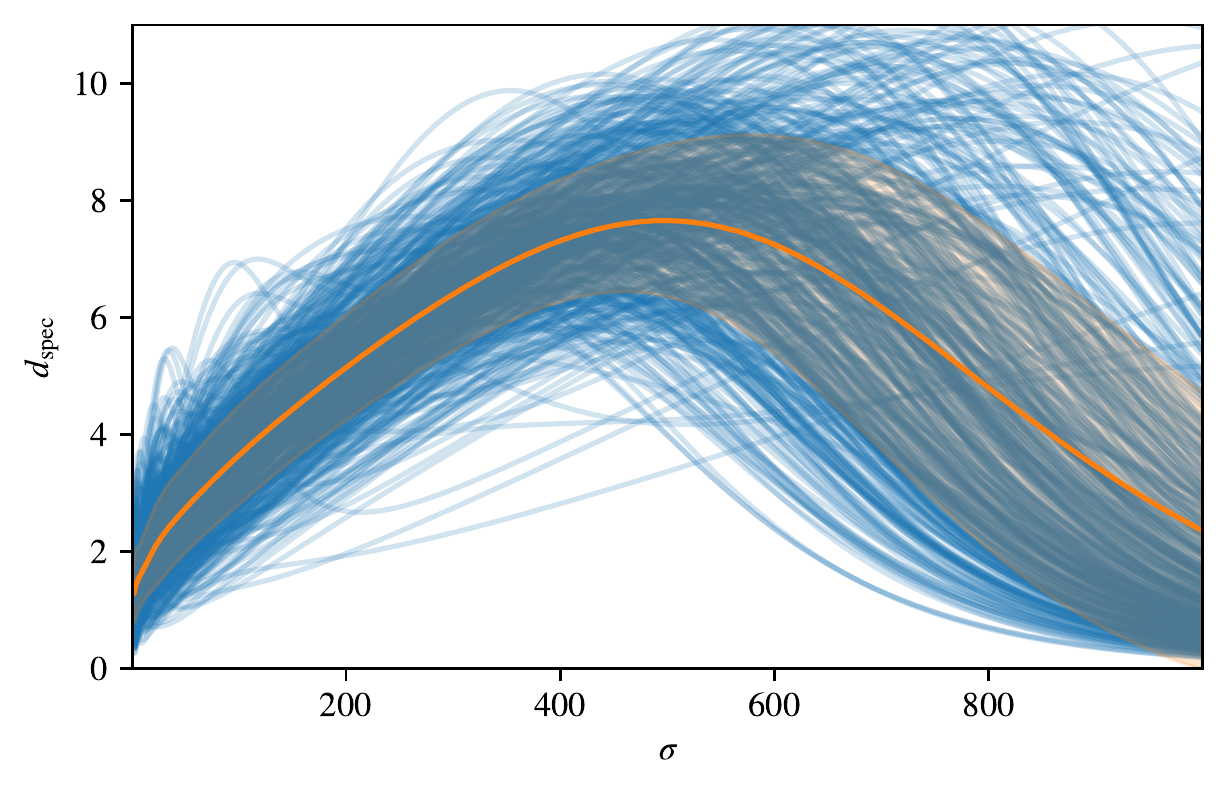}
\caption{\label{fig:transitive_percolations_1}
Spectral dimension for a causal set that arises from transitive percolation dynamics. The dynamics are modified such that a new element is only added if it is connected to at least one previously existing element. The causal set has $N=10^5$ nodes and was grown with $\beta_{tp}=5\cdot 10^{-6}$.
}
\end{figure}

So far, we have considered static networks. However, many real-world networks grow. Similarly, causal sets for quantum spacetime are expected to grow according to a dynamical growth rule. The similarity we have  observed between the spectral dimension in static causal sets and the spectral dimension in real-world networks motivates us to consider an existing model of a growing causal set and investigate its spectral dimension for the first time.
Transitive percolation causal sets \cite{Rideout:1999ub} are constructed by starting with one base element. Additional elements are added one by one. Every new element $x$ is causally related, $y_i \prec x$, to each of the existing elements $y_i$ with a probability $\beta_{tp}$.  A new element is only kept if it has at least one connection to the existing causal set, so that a totally connected graph is grown.  The resulting causal set's transitive reduction is computed to obtain the Hasse diagram\footnote{Notably transitive reduction has been used previously in the analysis of real-world networks \cite{transred}.}. We evaluate the spectral dimension on the corresponding network, cf.~Fig.~\ref{fig:transitive_percolations_1} for $\beta_{tp}=5 \cdot 10^{-6}$. It grows very slowly towards a maximum, before decaying without exhibiting a plateau.
In particular, the spectral dimension does not resemble that of a causal set based on a sprinkling endowed with a Lorentzian metric, nor that of real-world networks that are small world. This could have been expected based on the local nature of the growth process, which cannot capture the particular non-local structure necessary to build a discrete spacetime, or a small-world network. Accordingly, non-local growth dynamics are presumably necessary to grow either of these networks, see also \cite{Benincasa:2010ac}. Specifically, the rule to draw an edge between an existing and a newly added node cannot just depend on the two nodes in question, but must account for the neighbors of both nodes as well as the nodes none of the two nodes is connected to. Such a growth rule is computationally expensive, since it scales at least quadratically with the number of nodes.
\section{Conclusions}
We have explored the connection between networks that encode properties of quantum spacetime and networks that encode properties of real-world systems. We have focused on real-world networks which are embedded in a geometry, i.e., spatial networks. The embedding constrains the properties of these networks. To explore the impact of the embedding geometry, we have generated synthetic networks that are based on distances measured with the Euclidean, the Minkowski and a hybrid metric. We have used diffusion processes and the associated spectral dimension as an observable to compare the synthetic to the real-world networks. \\
First, we have shown how the dimensionality of the embedding geometry can approximately be recovered from a diffusion process on a road network. The corresponding synthetic network is based on a Euclidean metric, translating into a network with small degree.\\
Second, we have explored whether the dimensionality of the embedding geometry can approximately be recovered from networks with small-world property. This property is typically exhibited by networks which are designed to efficiently distribute information. They contain shortcuts that increase the degree and cannot be captured by a Euclidean metric. We have found that the spectral dimension of such real-world networks (including a representation of the internet as well as two neural networks) contains several classes of diffusion processes. More than one class is necessary to account for the heterogeneity of the network. Different starting points for a diffusion process result in different characteristic curves for the spectral dimension. Up to four different classes make up the spectral dimension in the networks we study. Remarkably, none of the classes resembles the characteristic curve of synthetic networks constructed with a Euclidean metric. Instead, one class resembles the  characteristic curve of synthetic spacetime networks that are constructed with a hybrid metric: at large enough Euclidean distance, their connectivity follows a Euclidean metric and encodes the dimensionality of the spacetime. At lower Euclidean distance, the connectivity of the synthetic network follows from the Minkowski metric. Such hybrid-signature networks do not (yet) play a significant role in quantum-gravity research, and have only been used as a regularization for causal sets. Our results suggest that such hybrid-signature networks could instead become important to describe real-world networks that exhibit the small-world property.\\
As an aside, we note that causal sets for Minkowski spacetime exhibit a characteristic of small-world networks in dimensions lower than four: the average network distance between any two nodes grows logarithmically with network size. In four dimensions, we find a constant average network distance and in higher dimensions we find a decreasing average distance. This might provide a new starting point to answer why four should emerge as the expected spacetime dimensionality from the path integral over causal sets.
In the future, importing further geometric notions, e.g., the Olivier curvature, from quantum gravity \cite{Klitgaard:2017ebu,Klitgaard:2018snm,vanderHoorn:2020rvd,vanderHoorn:2020rnz} could be of interest for real-world networks. Similarly, properties of Lorentzian geometry that can be extracted from a causal set, see \cite{Surya:2019ndm} for an overview, could also be measured for real-world networks to further determine similarities between those a priori rather distinct types of networks. This could provide information about whether small-world networks can be embedded in spacetimes with Minkowski or hybrid metric.
\begin{acknowledgements}
M.~P.~thanks S.~Guldner and S.~Siehl for valuable discussions on brain data.
This work was supported by VILLUM fonden under grant no.~29405. M.~P.~is  supported by  a  scholarship  of  the  German  Academic  Scholarship Foundation.
\end{acknowledgements}
\appendix
\section{Dimension-dependent scaling of the average shortest distance}
\label{app:scaling}
\begin{figure}
\includegraphics[width=\columnwidth]{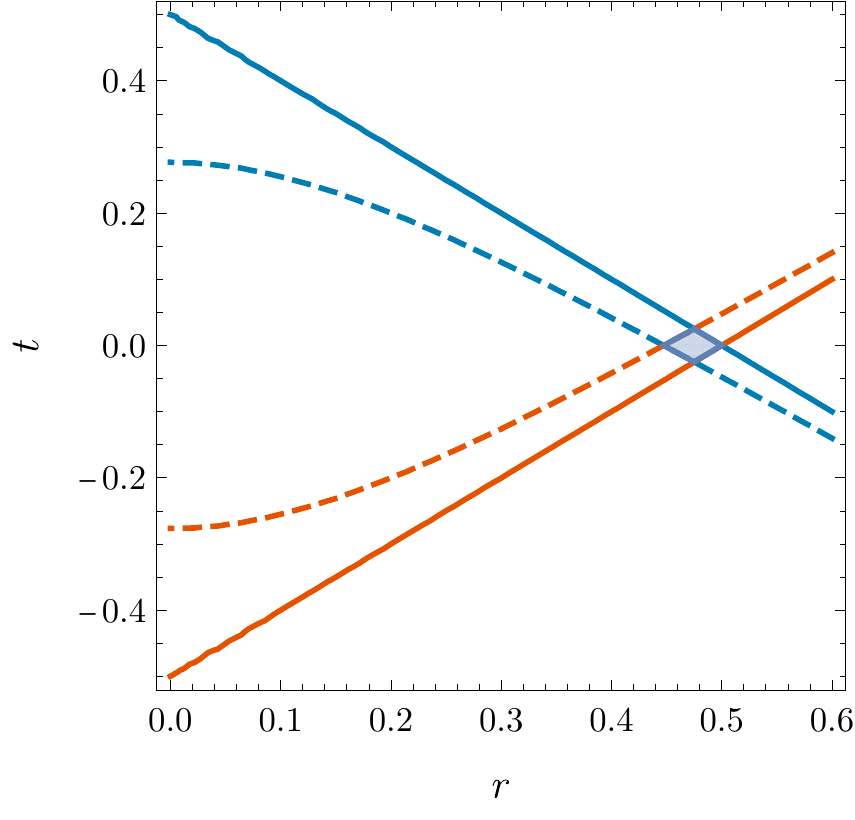}
\caption{\label{fig:lightcone}
Causal structure for two points $p$ and $q$. The solid lines mark the lightcones, the dashed once mark the boundary of the volume that contains most points with a direct connection to $p$ and $q$ and lie at a fixed Minkowski distance to $p$ or $q$, respectively. The blue shaded area is the intersection for which we are computing the volume $V$.
}
\end{figure}

\begin{figure}
\includegraphics[width=\columnwidth]{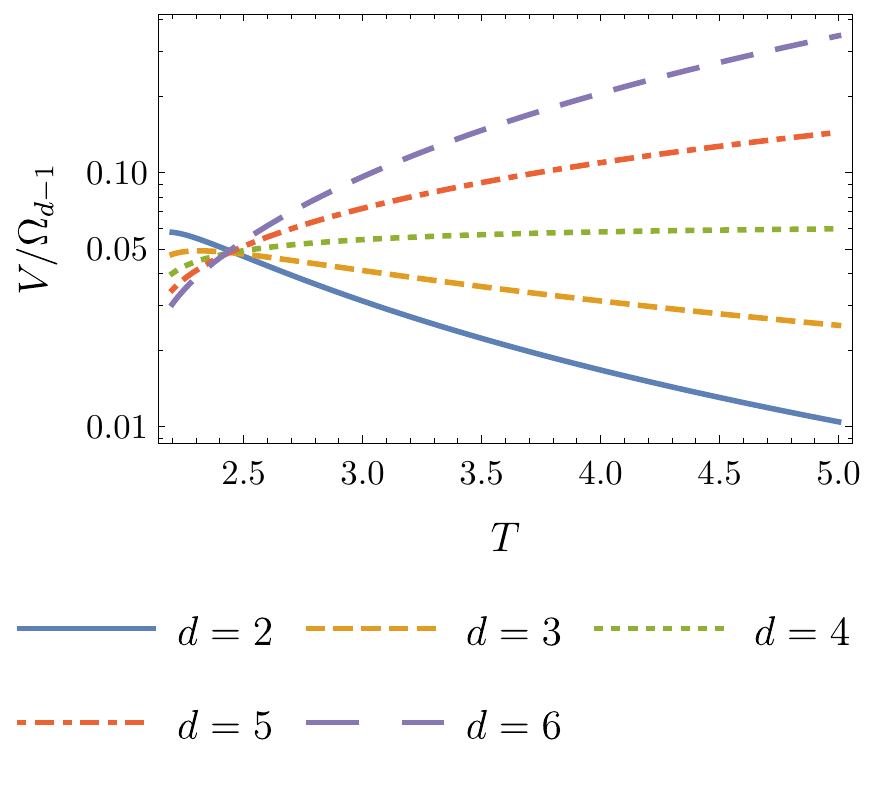}
\caption{\label{fig:scaling_volume_dimension}
Scaling of $V/\Omega_{d-1}$ by $T$ for varying dimension. Here we choose units such that $\rho = 1$.
}
\end{figure}

We investigate how the number of nodes relates to the average shortest distance between two random nodes for $d$-dimensional Minkowski space. Here, $d$ is the number of spacetime dimensions.
The relation is hard to extract analytically. To make progress we make two simplifications: we focus on the shortest non-trivial path length, i.e., on paths of length two and we fix the distance between the two nodes that we are considering. \\
In particular, we consider two nodes $p$ and $q$ placed at $(T/2, 0, \dots, 0)$ and $(-T/2, 0, \dots, 0)$.
We assume that the volume $V_\text{tot}$ enclosed by their respective backward- and forward-lightcones contains $N$ nodes, such that the density of nodes is $\rho = N/V_\text{tot}$.
The nodes $p$ and $q$ are connected by many chains. These chains feature a varying number of intermediate nodes.
We explore how frequent chains with only one intermediate node are.
Such an intermediate node is directly connected to both $q$ and $p$.
The large majority of nodes directly related to $q$ lies in a volume that is bounded by the light-cone emanating from $q$ and the hyperboloid given by \cite{Surya:2019ndm}
\be
\label{eq:app_hyperbolid}
-(t + T/2)^2 + r^2 = - \rho^{-\frac{2}{d}}
.\ee
The exponent on the right-hand side follows from dimensional considerations.
A similar volume exists for $p$. We want to  identify the volume that hosts nodes directly connected to both, $q$ and $p$. Such intermediate nodes lie in the intersection $V$ between these two volumes, cf. Fig.~\ref{fig:lightcone}.
For large $T$ the curvature of the surface \eqref{eq:app_hyperbolid} is negligible.
We hence expand the expression for $r(t)$ resulting from \eqref{eq:app_hyperbolid} to first order in $t$.
We can then compute $V$ as
\be
V = \Omega_{d-1} \int^{\frac{\rho^{-2/d}}{2T}}_{0} \dd{t} \int^{\frac{T}{2}-t}_{t_\text{hyp}} \dd{r} r^{d-2}
\ee
with
\be
t_\text{hyp} = \frac{T \left(\frac{T}{2} + t\right)-4 \rho^{-2/d} }{\sqrt{T^2 - 4 \rho^{-2/d}}}
.\ee
Here $\Omega_{d-1}$ is the integral over the $d-1$ dimensional unit sphere. In Fig.~\ref{fig:scaling_volume_dimension} we plot the resulting expression for varying dimension.
As apparent from this plot for $d=4$ the resulting volume remains nearly constant. Indeed, expanding $V$ for large $T$ yields $V = \frac{\pi}{4 \rho} + \order{\frac{1}{T^2}}$.
For smaller dimensions $V$ tends to zero, for larger dimensions $V$ diverges.
We assume that $p$ and $q$ are connected by only one intermediate node, if at least one node falls inside the volume $V$.
The fraction $V/V_\text{tot}$ gives the probability of placing a node inside $V$.
The probability for none of $N$ nodes to fall inside the volume $V$ is
\be
p_\text{no-conn} \equiv 1 - p_\text{conn} = ( 1- V/V_\text{tot})^N
.\ee
The probability $p_\text{conn}$ approximates the probability of $p$ and $q$ being connected with only one intermediate node.
In the limit $N \to \infty, \rho = N/V_\text{tot} = \text{const}$ we obtain $p_\text{conn} = 0$ for $d<4$, $p_\text{conn} = 1$ for $d>4$, and a finite value
$ 0 < p_\text{conn}
< 1$ for $d=4$.
This qualitatively agrees with our numerical findings reported in Sec.~\ref{sec:causal_sets_spectral_dimension}.
\section{Method and data sources}
In this appendix and Tab.~\ref{tab:data_sources} we describe our implementation to obtain the spectral dimension, the data sources that we consider and various data preparation steps.
We compute the spectral dimension by numerically performing a diffusion process.
For handling the graph data structure we use the SNAP library \cite{leskovec2016snap}.
We propagate probabilities.
At each time step $\sigma$, we store the probability $P(x, x', \sigma)$ for every node $x$ in the network.
We update probabilities by setting $P(x, x', \sigma + 1) = (1-\delta) P(x, x', \sigma)$ and $P(y, x', \sigma + 1) = \delta\, w(x,y) P(x, x', \sigma)$, where $y$ is a neighbor of $x$ and $w(x,y)$ is the weight for node $y$. In the unweighted case the weight is given by the inverse degree of $x$.
We repeat this update for every node $x$ in the network and sum the resulting probabilities at each node.
This yields the probabilities $P(x, x', \sigma + 1)$.
We extract $P(x', x', \sigma)$ for every $\sigma$ and compute the spectral dimension.
For the two road networks that we study, we utilize the following data sources:
\begin{itemize}
\item Data for the graph for the Pennsylvania street network is taken from the SNAP dataset collection \cite{leskovec2009community}.
\item The Europe roadnet dataset originally was compiled for the 10th DIMACS challenge \cite{dimacs_challenge} from OpenStreetMap data. As this graph contains many chains of degree-two nodes representing streets without intersections, we post-processed the graph in the following way:
\begin{enumerate}
\item We loop over all nodes.
\item If a node $i$ has degree two and any of its neighbors $j$ has degree two, then we first check that node $i$ and its two neighbors do not form a triangle. If they do not form a triangle we collapse nodes $i$ and $j$ into one node by removing node $j$ and and turning all neighbors of $j$ into neighbors of $i$.
\item We repeat step (2) until $i$ has no further neighbors to remove.
\end{enumerate}
This procedure contracts chains of degree two nodes into one edge representing the corresponding road. We have confirmed that we obtain similar results for random walks on the original graph with a slightly reduced spectral dimension.
\end{itemize}
The brain networks are extracted from the following sources:
\begin{itemize}
\item The Drosophila connectome is  taken from Ref.~\cite{xu2020connectome}. We use version 1.2 of the connectome. We have tested both, a weighted and an unweighted version of the resulting network. The results are similar.
\item The mouse network is obtained by taking the correlation matrix obtained in Ref.~\cite{knox2018high}, that measures the correlations between various voxels (three dimensional volumes) in a mouse brain. We model each voxel by a graph node. We then introduce a weighted edge between two nodes if they are linked by a correlation that is larger than the cutoff $ 5 \cdot 10^{-3}$.
\end{itemize}
For the internet graph, we use a CAIDA IPv4 topology dataset \cite{caida_dataset}, that represents independent network entities in the world wide web. In particular, we use the data from Feb.~28th, 2020.
\begin{table*}
\centering
\begin{ruledtabular}
\begin{tabular}{lrrp{8cm}r}
Graph & Nodes &  Edges & Comment & Source \\
\hline
Internet &     33304 &  69442 & CAIDA dataset from Feb. 28th 2020 & \cite{caida_dataset} \\
Drosophila &     21739 &  2897925 & Connectome of the Adult Drosophila Central Brain & \cite{xu2020connectome}\\
Mouse &     212894 & 143220733 & Voxel correlations from Allen Mouse Brain Connectivity Atlas with cutoff $r = 5 \cdot 10^{-3}$ & \cite{knox2018high} \\
Roadnet Pennsylvania &    1088092 & 1541898 & Network of Streets in Pensylvania & \cite{leskovec2009community} \\
Roadnet Europe &     16664809 & 19807451 & Network of Roads in Europe - reduced (see text) & \cite{dimacs_challenge}\\
\end{tabular}
\end{ruledtabular}
\caption{\label{tab:data_sources}
We list the real-world networks that we consider with the corresponding data sources. For more information see the main text.
}
\end{table*}
\bibliography{references,references_additional}
\end{document}